\documentclass[]{aa}

\usepackage{txfonts}
\usepackage{graphicx,curves,color,rotating,natbib}
\bibpunct{(}{)}{;}{a}{}{,}

\begin{document}

\title{The spatial distribution of carbon dust in the early solar nebula and  the carbon content of planetesimals}


\author{Hans-Peter Gail\inst{1}  \and Mario Trieloff\inst{2,3}
}

\institute{
Institut f\"ur Theoretische Astrophysik, Zentrum f\"ur Astronomie,
           Universit\"at Heidelberg, 
           Albert-Ueberle-Str. 2,
           69120 Heidelberg, Germany 
\and
Institut f\"ur Geowissenschaften, Universit\"at Heidelberg, Im Neuenheimer
           Feld 236, 69120 Heidelberg, Germany
\and
Klaus-Tschira-Labor f\"ur Kosmochemie, Universit\"at Heidelberg, Im Neuenheimer Feld 236, 69120 Heidelberg, Germany
}

\offprints{\tt gail@uni-heidelberg.de}

\date{Received date ; accepted date}

\abstract
{A high fraction of carbon bound in solid carbonaceous material is observed to exist in bodies formed in the cold outskirts of the solar nebula, while bodies in the terrestrial planets region contain only very small mass fractions of carbon. Most of the solid carbon component is lost and converted to CO during the spiral-in of matter as the sun accretes matter from the solar nebula.}
{We study the fate of the carbonaceous material that entered the protosolar disk by comparing the initial carbon abundance in primitive solar system material and the abundance
of residual carbon in planetesimals and planets in the asteroid belt and the terrestrial planet region.
}
{From observational data on the composition of the dust component in comets and of  interplanetary dust particles, and from published data on pyrolysis experiments, we  construct a model for the composition of the pristine carbonaceous material that enters the inner parts of the solar nebula during the course of the build-up of the proto-sun by accreting matter from the protostellar disk. Based on a one-zone evolution model of the solar nebula we study the pyrolysis of the refractory and volatile organic component and the concomitant release of high-molecular-weight hydrocarbons under quiescent conditions of disk evolution where matter migrates into the central parts of the solar nebula. We also study the decomposition and oxidation of the carbonaceous material during violent flash heating events, which are thought to be responsible for the formation of chondrules, by calculation of pyrolysis and oxidation of the carbonaceous material in temperature spikes that are modelled according to cosmochemical models for the temperature history of chondrules. 
}
{It is found that the complex hydrocarbon components of the carbonaceous material are removed from the disk matter in the temperature range between 250 and 400\,K, but that the amorphous carbon component survives to temperatures of 1\,200 K. Without efficient carbon destruction during flash-heating associated with chondrule formation the carbon abundance of terrestrial planets, except for Mercury, would be several percent and not as low as it is found in cosmochemical studies. Chondrule formation seems to be a process that is crucial for the carbon-poor composition of the material of terrestrial planets. 
}
{%
}


\titlerunning{Carbon content of planetesimals}
\authorrunning{H.-P. Gail \& M. Trieloff}

\maketitle

\section{Introduction}
\label{SectIntro}

Models of the composition of interstellar dust require that a fraction of about 50\% of the carbon content of the interstellar matter is bound in a solid phase \citep[e.g.][]{Mat77,Zub04}, most likely an amorphous form of solid carbon, and additionally up to 10\% of the carbon is in large PAHs. A much more detailed model for the volatile carbonaceous component of the ISM dust is developed in \citet{Jon13,Jon14,Jon16}, \citet{Koe15}, and \citet{Ysa15}. An overview on this model is given in \citet{Jon17}. The solid material of accretion disks around newly formed stars is derived from such material, though probably somewhat modified in the parent molecular cloud and during the collapse phase to the protostar before it is  added to the disk. The dust component of an accretion disk is therefore expected to contain a considerable fraction of carbonaceous material. By carbonaceous material we mean any solid material chemically composed largely of C alone or in combination with the elements H, N, O, S, but not compounds with the typical rock forming metals. 

It is expected, thus, that in particular the material from the cool outer regions of the solar nebula which has not yet been subject to significant chemical and physical processing in the warm inner disk regions should be rich in solid carbonaceous materials. This expectation is corroborated by studies of infrared spectra of cometary comae, by direct studies of cometary material of comet Halley and comet Wild~2, and by studies of interplanetary dust particles (IDPs) and micrometeorites which all find high abundances of carbonaceous material \citep[see the reviews of][]{Woo02,Woo08,Ehr04}, and by the fact that carbonaceous chondrites --- the most pristine meteoritic material --- contains some mass-percent of carbonaceous material \citep{Gra03}. 

Contrary to this, bodies of the inner solar system ($\lesssim3$\,AU) --- the distance range where terrestrial planets and the parent bodies of the ordinary chondrites are found --- are almost void of carbon. Somehow the initially abundant condensed carbonaceous material disappeared from this region. Either the carbon is converted into gaseous species that are accreted by the sun and not incorporated into the forming bodies in this region (or only in tiny quantities), or it is first incorporated into the planetesimals but then converted to volatile species which escape from the bodies and finally also end-up in the sun, or, in differentiated bodies, it went into the iron core.

The problem of how the solid phases of carbon nearly completely disappear from the disk material in the source region of the terrestrial planets seems not to have attracted much interest. Because in the strongly hydrogen-dominated elemental composition of the solar nebula in chemical equilibrium almost all carbon would be present as gaseous CO or CH$_4$, it seems at first glance natural that essentially no carbon is built into the terrestrial planets because they form from the solids in the accretion disk.  Studies of the chemistry of the accretion disk show unequivocally, however, that the chemistry in the solar nebula and in particular the carbon chemistry is far from chemical equilibrium \citep[e.g.][]{Hen13,Wal14}, and the combustion of carbonaceous material requires high temperatures because of high activation energies of key reaction steps, as is known from flame chemistry \citep[see][]{War06}. The carbonaceous material that is incorporated into the solar nebula during its formation therefore is not easily converted into gaseous products which do not contribute to the planetary material in  the range of terrestrial planets.

The only work so far that has addressed the problem of how the solid carbon phases present in the cool outer regions of the solar nebula could be converted into gaseous species that are not accreted during planetesimal and planet forming processes seems to be the work of \citet{Lee10}. This work discusses photo-processes in the disk atmosphere that ultimately result in oxidation of carbonaceous materials.

We intend to explore in this paper the possibility that the carbonaceous material is converted in the bulk disk close to its mid-plane by pyrolysis in and by chemical reactions. Because of the much higher fraction of the disk mass involved in reactions operating in this region of the disk they may be more efficient in destroying the carbonaceous material.

The plan of our paper is as follows: in Sect. 2 we discuss the empirical data gained from comets and interplanetary dust particles (IDP) on the composition and morphology of solid carbonaceous material in the cold outer parts of the solar nebula. In Sect. 3 we give a sketch of the processes which may be responsible for destruction of the carbonaceous material. In Sect. 4 we outline a simplified model composition of the volatile and refractory components of the carbonaceous material and discuss in Sect. 5 the pyrolysis and oxidation of such material and the subsequent gas phase reactions. This chemical model is applied in Sect. 6 to a quiescent solar nebula with mass accretion. In Sect. 7 we consider the role that chondrule formation plays for the fate of the carbonaceous material in the solar nebula. In Sect. 8 we discuss the consequences of our results for the carbon content of the parent bodies of meteorites and for planets. Section 9 contains some final remarks. In three appendices we briefly discusse the pyrolysis reactions of the carbonaceous material, the gas phase reactions following the destruction of carbonaceous material, and present results of a sample calculation.

\section{%
Carbonaceous material in the solar nebula}
\label{SectCarbon}

The most direct information that we presently have on the pristine composition is that from comets because in that case it is clear that the material is unmodified material from the accretion disk \citep{Mum11}. Another source of information is the material found in anhydrous chondritic IDPs. This material is generally assumed to be of cometary origin and is likely to have been only slightly modified during its passage through the atmosphere. 

Chondrites also contain sometimes a very pristine material which might be traced back to the parent molecular cloud \citep[e.g.\ ][]{Ale07}, but generally the material of chondrites has been chemically processed by parent body processes and information from that source is only partly suited for our purpose. Also the very detailed model on the structure of the interstellar carbonaceous dust component developed during the last years \citep[see][ for a detailed review]{Jon16} is only of limited use for our purpose. This material is modified in the hot molecular core during the collapse of a molecular cloud core and during passage through the accretion shock until it is added to the accretion disk. Subsequently it is subject to chemical and thermal processing during its residence in the disk until it is built into planetesimals.

\subsection{%
Analysis of cometary material}  
\label{SectAnalCom}

During the apparition of comet Halley in 1986 an armada of spacecrafts were sent to a rendezvous with the comet's nucleus. Three of them succeeded in a close flyby at the nucleus and investigated the elemental, isotopic, and chemical composition of dust grains by their on-board dust-impact time-of-flight mass spectrometers \citep[cf., e.g., ][ for a review]{Jes99}. Over 5\,000 dust grains could be analysed. 

The composition of carbonaceous dust particles from comet Halley is discussed in \citet{Fom94b} and  reviewed by \citet{Fom97,Fom99}. According to this, the  carbon found in solid materials (dust particles) is distributed over a number of phases with different compositions. 

1.~Most of the carbonaceous material, a fraction of $\sim0.6$ by total number of carbon atoms in solid phases, is found in what has been called \emph{complex organic matter}. This is a complex refractory material containing C, H, N, O, and S with an average elemental composition C$_{100}$H$_{115}$N$_9$O$_{65}$S \citep{Fom94a}. The oxygen in this material may be found in hydroxyl, carbonyl, ether, or carboxyl groups. The nitrogen may be found in amine or nitrile groups. The exact composition of this mixture of a large number of compounds, however, cannot be determined from the kind of data that was acquired, but the structure of the insoluble organic matter in meteorites \citep[see][ for an attempt to characterize this]{Der10} may give an impression how the chemical structure of the material could look like. It may be that this material resembles the hydrocarbons of the next group with the oxygen-and nitrogen-bearing groups added to them. 

2.~The second most abundant component is a phase composed of hydrocarbons, containing a fraction of $\sim0.1$ of all carbon in solids \citep{Fom94b}. It contains compounds of C and H and is composed of (1) a component formed by hydrogenated amorphous carbon or polycyclic hydrocarbons with an average H/C ratio of 0.14, (2) a mixture of PAHs with an average H/C ratio of 0.5, (3) a component formed by a mixture of PAHs and highly branched aliphatic hydrocarbons with an average H/C ratio of 1.0, and (4) a component formed from highly branched aliphatic hydrocarbons with a typical H/C ratio of about 2.0.

3.~An almost pure carbon component contains a fraction of $\sim0.10$ of carbon in solids \citep{Fom94b}. Other elements are contained at a level of at most 1\%. Some of such particles have high $^{12}$C/$^{13}$C isotopic ratios, i.e., the presolar carbon grains are members of this phase. The carbon particles are generally very small \citep[$<0.05\,\mu$m, see][]{Fom99}.

4.~The remaining fraction of the carbon was found to be bound in some kind of carbon suboxide \citep[$\sim0.01$, ][]{Fom94b} and cyanopolyynes \citep[$\sim0.01$, ][]{Fom94b}. We do not consider these low-abundance components.
   
These carbonaceous components may, at least in part, be inherited from the ISM and/or the parent molecular cloud. The expected structure and composition of the carbonaceous component of the interstellar dust is discussed in detail in \citet{Jon13,Jon14}, \citet{Jon16}, \citet{Koe15}, and \citet{Ysa15} where a model is built based on reasonable dust analogues and on laboratory work on their properties and on their processing under interstellar conditions. The expected similarity in particular of the complex organic matter of Halley, the insoluble organic matter in carbonaceous chondrites, and interstellar carbonaceous dust suggest that they are closely related materials. Possibly part or even all of the complex organic matter of Halley and of the insoluble organic matter in carbonaceous chondrites is material from the parent molecular cloud \citep[see, e.g., ][ for a discussion]{Ale07, Ale11}. It would be interesting and useful to study the compositional relation between the carbonaceous components in pristine ISM dust and the dust from the solar nebula, but this is not the topic of this paper.
Nevertheless, a rough constraint on the fraction of surviving interstellar grains can be obtained from the abundance of presolar or circumstellar grains found in primitive meteorites or interplanetary dust particles. 
In primitive meteorites and also in IDPs which belong to the most primitive matter considered in our paper, certain refractory presolar grains have quite low abundances, e.g., SiC ranges from a few to c. 100 ppm, graphite c. 5--10 ppm, only non-carbonaceous silicates have average abundances of 400 ppm in IDPs, with exceptional cases reaching a few 1000 ppm. This means that even the fraction of surviving refractory presolar material seems at best of the order of a few permil
\citep[see., e.g., ][]{McS10}.  As these grains are more resistant against chemical destruction than more volatile components, a few permil seems to be an upper limit for the fraction of surviving volatile components.
Moreover, the mere presence of bulk isotopic anomalies in D/H does not imply that the carrier phases were not chemically modified. It just means that their isotopic ratio did not equilibrate completely with other material having the bulk solar system value. Hence, we share the view of many meteoriticists that most of the solar system material was homogenised in the protosolar disc or at least chemically modified.

The abundance of carbon relative to Si and its fraction bound in carbonaceous solid material was studied by \citet{Gei87}. According to this the C/Si abundance ratio is close to solar. The solar abundance of carbon had to be revised downwards in the meantime \citep{Asp09} but in view of the considerable uncertainties of abundance determinations for Halley the basic statements remain unchanged. In the material released by Halley a fraction of $\sim0.47$ of the carbon is bound in the solid carbonaceous phases while the remaining fraction is found in the released gas \citep{Gei87}. It has to be noted that the dust grains from Halley were analysed a few hours after their release when they are already heated to a high temperature  by illumination with sunlight for some hours such that the more volatile condensed phases evaporate. That some of the less-refractory carbonaceous material is desorbed before measurement by the space probes is suggested by the results for the distribution of gas species in Halley's coma which seem to require that there is an extended source of carbon in the coma \citep{Ebe87,Cot08}. 
 
Attempts have been undertaken to identify PAHs in the coma spectrum of Halley and it has been claimed that pyrene (C$_{16}$H$_{10}$) and anthracene (C$_{14}$H$_{10}$) are detected by emission bands in the UV \citep{Cla04,Cla08}. The results, however, are inconclusive. Such PAHs were identified, however, in carbonaceous material from dust particles collected in the tail of Wild~2 \citep{Cle10}, cf. also the review of \citet{Li09}. This shows that PAHs are indeed present in cometary material. 

The carbonaceous material seems to form in most cases a coating on a chondritic core \citep{Kis87}.

The results derived from observations of Halley are much more conclusive than inferences from of the carbon content of Wild 2 particles \citep{Bro06}, because these particles were captured at high speed in an aerogel und may have experienced significant loss of volatiles incl. carbon during the collection process. 

\subsection{%
Laboratory analysis of interplanetary dust particles}
\label{SectAnalIDP}

Interplanetary dust particles are typically 10\,$\mu$m sized aggregates of a large number of mineral grains ($>10^4$) and carbonaceous material. They show fluffy structures of very small grains with diameters between 1 and 150 nm composed of glass, crystalline mineral grains, and Fe,Ni-metal and Fe-rich sulphide crystals embedded in glass, with abundant carbonaceous material apparently acting as a glue. The porosity of these aggregates is very high, between 0.7 and 0.94 \citep{Fly13}.

The IDPs are enriched by carbon and other moderately volatile elements by factors around two to three compared to CI chondrites \citep{Fly96}. Some IDPs show strongly anomalous D/H isotopic ratios suggesting that they incorporated molecular cloud material pre-dating the solar nebula \citep[e.g.][]{Kel00}.  After their formation they obviously have never been heated to temperatures sufficient for evaporation of the moderately volatile elements \citep{Fly96} and the subclass of anhydrous chondritic porous interplanetary dust particles (CP-IDP) therefore apparently represents the most primitive available material from the early days of solar system formation \citep[cf. the review of ][]{Mes06}. They are most likely derived from comets from the Kuiper-belt region such that their composition probably can be taken as being close to the initial composition of solar nebula solids. The presence of some fraction of crystalline material \citep{Bra94} means, however, that they contain also some fraction of material from the inner parts of the solar nebula, such that their composition is not completely identical with the composition of the material that rained down from the parent molecular cloud of the solar system. 

The carbon content of the IDPs is highly variable and amounts on average to 12 weight-percent \citep{Thomas94} and may be as high as 46\% \citep{Tho93}. The volume fraction of the carbonaceous material is between a few percent and up to 90\% \citep{Fly03b}. The spatial distribution of the carbonaceous material within IDPs was investigated by scanning transmission X-ray microscope (STXM) \citep[cf.][]{Fly03,Fly04}. The carbon is observed to form three morphologically different structures in IDPs:

\begin{itemize}
\item Thin coatings ($\sim100$ nm) on individual grains. The individual grains are generally not in contact with each other but separated by such a coating. A very instructive picture is shown in  \citet[ Fig. 2]{Fly13}

\item Discrete sub-$\mu$m to $\mu$m carbonaceous regions.

\item Thick ($\sim0.5\,\mu$m) coatings on aggregates of grains, acting as a ``glue'' which appears to hold the subunits together. 

\end{itemize}
As \citet{Fly03} noted: "The observation that carbon occurs as coatings on grains, apparently being the material that holds the subunits of the IDPs together, indicates that this carbon existed at the time when grains were being assembled into dust-size particles." The morphology of the carbonaceous material in IDPs tells us therefore the initial form of appearance of solid carbon in the solar nebula.
   
The composition of the carbonaceous material was investigated by X-ray Absorption Near Edge Structure (XANES) spectroscopy and by Fourier transform infrared spectroscopy (FTIR) \citep[e.g.][]{Fly03,Fly04,Fly13}. 

1. From XANES it is found that purely amorphous carbon is not very abundant. A  characteristic feature for amorphous carbon in the K-edge XANES spectrum of C was found for only one of the twelve investigated particles \citep{Fly03}. 

2. Most carbon is present as some kind of organic material, witnessed by a high oxygen content of the carbonaceous material \citep{Fly03}. This was detected by XANES of the K-edge of C which revealed strong absorption features that are identified with abundant C=O bonding and with C=C bonding in ring structures of the C atoms  \citep{Fly03,Fly13}. The absence of a feature characteristic for graphitic carbon indicates that this form of condensed carbon is (nearly) absent \citep{Fly04}.

3. By FTIR spectroscopy of the 3\,$\mu$m regime it was found that from the 19 IDPs investigated 17 showed strong C-H stretching vibrations of aliphatic hydrocarbons, associated with CH$_2$ and CH$_3$ groups \citep{Fly03}. The relative strength of the features is highly variable for different particles indicating a variable mixture of different aliphatic hydrocarbons. The average value of the CH$_2$/CH$_3$ ratio is $\sim2.46$ for anhydrous IDPs \citep{Fly03}. Since for simple aliphatic hydrocarbons the CH$_3$ groups terminate the chains of CH$_2$ groups, the CH$_3$/CH$_2$ ratio is a measure for the chain length of the aliphatic hydrocarbons, but this depends strongly on the unknown branching ratio and is not easily converted into a chain lenght.

The mass-fraction of aliphatic hydrocarbons in IDPs was estimated in \citep{Fly04} to be approximately 1 to 3 weight-percent. Since total carbon abundance is on average 12 weight-percent, the aliphatic hydrocarbons comprise a fraction of $\sim0.1\dots0.2$ of the carbon.  

4. It is reported that only a few of the investigated IDPs show C-H stretching vibrations characteristic for aromatically bound C, and these are found to be weak. Thus the fraction of aromatically carbon in IDPs seems to be low \citep{Fly03}.
 
5. For one IDP \citet{Fly08} succeeded to determine the composition of the organic coating of one of the silicate grains. By XANES an average composition of C$_{100}$N$_{10}$O$_{50}$ was derived.
 
These findings with respect to the composition of the carbonaceous material in IDPs are essentially in accord with the results found from analysis of Halley dust. Since the anhydrous IDPs originate almost certainly from a number of different comets with different formation regions in the solar nebula, one has to conclude that the refractory condensed phases of carbon have a rather uniform composition in the outer parts of the solar nebula, despite of the significant particle-to-particle variations. Except for the small fraction of pure carbon grains, most of the refractory carbon phases are found in aggregate dust particles as coatings on mineral dust grains.

\section{%
Metamorphosis of the carbonaceous material in the solar nebula}
\label{SectProcDu}

Our main interest is focussed on the carbon content of the central parts of the Solar Nebula where the terrestrial planets and parent bodies of the ordinary chondrites are formed. During the early evolutionary phase of an accretion disk like the Solar Nebula one has a general net flow of matter through the inner disk toward the proto-sun, the accretion flow. The material which is accreted by the still growing sun from the inner edge of the disk is continuously replenished by material initially located farther out in the disk. The residence time of the material in the inner disk is estimated to be less than 10$^5$ years if one assumes that the mass contained in the inner disk region is of the order of 10$^{-3}$ M$_{\sun}$ and the accretion rate is a few times $10^{-8}$\,M$_{\sun}$\,yr$^{-1}$ \citep[which may be typical values during the initial evolution period, ][]{Man12,DaR14}. Planetesimal  formation commences at probably a few times 10$^5$ years after disk formation in the innermost part of the solar nebula and at about $1\dots2\times10^6$ years at the distances of the present Asteroid Belt. At this instant the material in the inner disk region is no more identical with the  material that was deposited in this region during the initial formation period of the accretion disk. Instead, it is  material that originates from cold disk regions and migrated inwards during the course of the accretion process.    

\begin{figure*}

\includegraphics[width=\hsize]{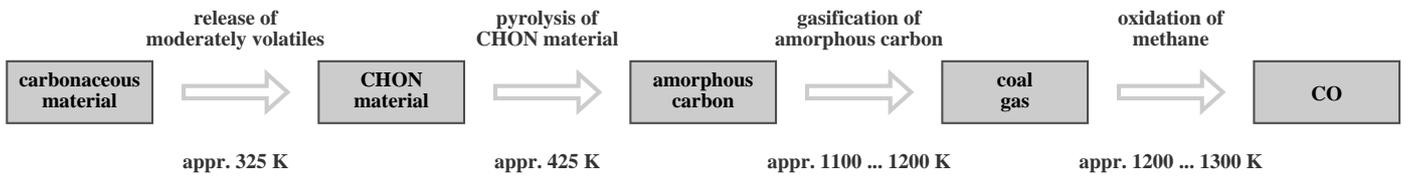}

\caption{Processes responsible for the conversion of the carbonaceous material observed for cool disk regions into gaseous components and their final oxidation to CO.}
\label{FigMetamorCar}

\end{figure*}

In the cold outer regions of the accretion disk where water ice is frozen the composition of the refractory solids is maintained as it is since the time when the matter rained down to the accretion disk from the parent molecular cloud of the sun. Temperatures are too low in this region for any physical or chemical modification of the minerals. It is not clear whether this also holds for the refractory carbonaceous components. The kind of material found in IDPs shows a composition which seems to be different from what is assumed for the composition of the interstellar carbon dust. The corresponding compositional changes ought to occur either in the parent molecular cloud or in the accretion disk itself. The high D/H isotopic ratios found in the carbonaceous material of some IDPs are interpreted as pointing to an origin of at least part of the carbonaceous material in the extremely cold environment of a giant molecular cloud \citep[e.g.][]{Ale11}. But if such material can be formed in the parent molecular cloud, it can very likely also be formed in the cold outer parts of the accretion disk, where similar conditions are encountered with respect to the chemistry acting as in molecular clouds. It has been argued \citep{Der10} that the insoluble organic matter in carbonaceous chondrites, which must be somehow derived from the carbonaceous matter in IDPs, is of Solar Nebular and not ISM origin.

For our purposes the question of the true origin of the carbonaceous material is not important. All what we need to know is the composition and morphology of the carbonaceous material as it migrates into the inner disk region. At this place we take recourse to what is learned from investigations of cometary particles and IDPs. These showed that the carbonaceous component of the dust aggregates is composed of four essentially different components:

\begin{enumerate}
\item Pure amorphous carbon grains.

\item An aliphatic hydrocarbon component.

\item An aromatically bound hydrocarbon component.

\item Refractory carbonaceous material with rather high contents of O, N, and S and with a not-well-defined composition.

\end{enumerate}
The second, third, and fourth component form a coating on the mineral grains. This mixture of carbon compounds defines the initial composition of that fraction of the carbon that enters  the inner part of the solar nebula in the form of refractory condensates. The remaining fraction of carbon is in volatile ices and in the gas phase (CO). More details will be specified later. 

During the evolution of the accretion disc there is a general average slow inflow of matter from cold outer regions toward the centre. The general temperature structure of the disk during the about first million year of its evolution is such that the temperature in the interior of the disk increases with decreasing distance from the proto-sun. If we refrain for the moment from the more complex local flows superposed on this long-term accretion flow, the matter is systematically transported by inflow into regions of increasingly higher temperature where chemical reactions are activated and finally reaction timescales become short enough for establishing chemical equilibrium. 

As long as the dust component of the disk matter is not separated from the gas and enclosed in planetesimals it is chemically coupled to the gas phase. The highly fluffy structure observed for IDPs means that the elementary building blocks of the dust aggregates are all in direct contact with the gas phase. The chemical compositions of the dust- and gas components evolve simultaneously during gradual heating caused by inward migration of the matter. They are coupled by solid-gas interface reactions. If planetesimals are formed from the solid component of disk matter in some region of the accretion disk, their initial composition corresponds to the local composition of the solids at that instant and that region. The subsequent evolution of the internal chemical composition of planetesimals and of the residual gaseous material in the disk follow different path's from that moment on. 

Our interest concentrates on the question how much of the initial refractory carbon dust material still exists in different zones of the solar nebula that can be incorporated into planetesimals and later into planetary bodies. To answer this we have to study the evolution of the initial material as it is heated up during inwards migration. The individual carbon compounds constituting the initial material differ in their volatility. The aliphatic hydrocarbons found in cometary material are moderately volatile. They are transferred to the gas phase at rather low temperature by evaporation. The small fraction of amorphous pure carbon is highly refractory. This component survives up to rather high temperatures \citep[$>1\,000$ K, see][]{Gai01,Gai02a} until it is destroyed by oxidation reactions and is finally converted into CO. The main part of the carbonaceous material is a refractory organic material probably similar to the material kerogen known from earth which evaporates and decomposes at several hundred K forming gaseous hydrocarbons and a residue resembling amorphous carbon.

The conversion of the refractory carbon into gaseous species enriches the gas phase with hydrocarbons. In chemical equilibrium these would be converted into CO and CO$_2$. Since the gas phase reactions for oxidation of the hydrocarbons involve relatively high activation energy barriers,\footnote{%
Required to dissociate H$_2$O to form reactive OH radicals in case of accretion discs, or to dissocate O$_2$ to form O radicals in case of terrestrial flames. The O and OH radicals attack hydrocarbons leading to formation of a C-O-bond and via a chain of follow-up reactions ultimately to CO molecules.
} the oxidation reactions require rather high temperatures to become efficient \citep[$\gtrsim1\,200$ K in the solar nebula, ][]{Gai02b}. From laboratory studies on combustion of gaseous fuels it is known that during the course of such processes complex organic compounds are to large extent first degraded into mainly C$_2$H$_2$ and CH$_4$ and the final oxidation then heavily relies on the hot gas-phase chemistry of such molecules \citep[e.g.][]{War06,Mar96}. Hence, there will be a certain zone in the accretion disk where the complex aliphatic and aromatic compounds are converted into simple low-molecular-weight hydrocarbons which finally are converted to CO in a hotter region.  

Hence, the main processes that have to be considered for studying the metamorphosis of carbon in the solar nebula are (cf. Fig.~\ref{FigMetamorCar}):
\begin{itemize}
\item Evaporation and condensation of the volatile components of the carbonaceous mixture. 
\smallskip

\item Pyrolysis of the refractory organic components.
\smallskip

\item Gasification of the residual coke by reaction with hot steam (so called town-gas reaction).
\smallskip

\item Final conversion of the hydrocarbons by further reaction with hot steam to CO.
\smallskip

\end{itemize}
The combination of these physico-chemical processes with the transport of material in the accretion disk and the gradual evolution of the disk properties determine the fraction of the total carbon existing as condensed carbonaceous material at any instant and location in the solar nebula and, in turn, determine the amount of carbonaceous material incorporated into planetesimals.

\section{Model for the dust component}

In order to determine the quantities of the different gaseous and condensed carbon bearing species in the solar nebula we have to model the chemical reaction processes between all carbon compounds in the gas phase, the exchange of material between the gaseous and condensed material, and the chemical surface reactions. Because of the complex structure and composition of the dust this requires some simplifications. In the following we introduce such a model for the carbonaceous material which is guided by what is known about IDPs and cometary dust and concentrates on the basic aspects which are likely to be important for modelling the distribution of carbon dust in the warm inner disk regions. 

\begin{table*}

\caption{%
Assumed composition of the carbonaceous material in cold disk regions. The fractions of carbon bound in the different components are given with respect to the total carbon bound in the carbonaceous material, which in turn is a fraction of 0.55 of the C element abundance. For the different components the gases assumed to be released by heating or oxidation and the corresponding approximate release temperatures (in K) and the relative proportions of the products are given. }

\begin{tabular}{llclclr}
\hline
\hline
\noalign{\smallskip}
          &           & Fraction $f$ &\multicolumn{2}{c}{Outgassing} & \multicolumn{2}{c}{Released} \\
Carbonaceous dust component & Structure &  of C & Process  & Temperature & Species & Percent \\
\noalign{\smallskip}
\hline
\noalign{\smallskip}
Pure carbon dust & Amorphous carbon & 0.1 & Oxidation  & 1\,100 & HCCO & 100  \\[.1cm]
Volatile organic material & Aliphatic compounds & 0.1 & Pyrolysis  & 325 & CH$_4$ & 50 \\
 & &  &  &  & C$_4$H$_{10}$ & 50 \\[.1cm]
Moderatly volatile organics & Aromatic compounds & 0.1 & Evaporation  & 425 & C$_{16}$H$_{10}$ & 100 \\[.1cm]
Refractory organic material & CHON & 0.6 & Pyrolysis  & 425  & CH$_4$ & 15 \\
 & & & &  & CO & 25 \\
 & & & &  & CO$_2$ &  20 \\
 & & & &  & C$_{16}$H$_{10}$ & 40\\[.1cm]
Others &  & 0.1 &   &   &  \\
\noalign{\smallskip}
\hline
\end{tabular}

\label{TabCarbComp}
\end{table*}

\subsection{Structure of dust particles}
\label{SectDuStru}

The observed IDPs are complex aggregates of thousands of tiny mineral dust grains with sizes $\lesssim0.1\,\mu$m, each of which is coated with carbonaceous material which seems to act as a glue for the aggregate (see Sect.~\ref{SectCarbon}). The dust material entering the outer edge of the computational domain is assumed to have the morphology as it is observed for IDPs. 

The porosity of the aggregates is very high with observed values in the range $0.7\dots0.94$ \citep{Fly13}. Since the aggregates were slightly under pressure by self gravity of the cometary parent body and, as a consequence, are slightly compacted during their residence in the comet \citep[e.g.][ Fig. 2]{Hen12}, it is likely that the highest porosities found for IDPs resemble their typical porosity before they were initially incorporated into the parent comet. At such high porosities the aggregates are mainly highly branched linear chains of particles. With respect to gas-grain collisions the surface of each grain of such a highly fluffy grain aggregate is essentially in direct contact with the gas phase into which the aggregate is suspended. Inside such a fluffy aggregate, the probability that a gas-phase particle experiences multiple collisions with the aggregate's constituent grains is negligibly small.

For calculating evaporation/condensation processes or chemical reactions between gas phase species and surface atoms this allows us to introduce the simplification that for the purposes of modelling the chemistry (and only for this purpose) each grain of an aggregate is replaced by a fictitious single isolated grain that interacts with the gas phase. Each of this particle is assumed to be initially coated with a mantle of carbonaceous material as it seems to be the case for the real IDPs. Any previously existing ice coating has already gone at the temperatures of interest ($>200$ K). According to \citet{Law92} neither pure silicate nor pure CHON particles seem to be detected in the dust from comet Halley. For simplicity we assume additionally that all the carrier grains of the carbonaceous mantles are spheres and all have the same radius which is arbitrarily set to 50\,nm.   

In principle the dust aggregates may drift with respect to the gas. The velocity calculated for the whole aggregates then have to be prescribed for each of its grain members. This becomes important at the stage of evolution where planetesimals are formed. Since the main subject of the present paper is the carbon chemistry in the solar nebula during the period preceding planetesimal formation, however, we presently refrain from considering dust drift in our model calculation.

\subsection{Model composition of carbonaceous material}

We assume that the composition of the carbonaceous material found in anhydrous chondritic IDPs and observed for comet Halley is representative for the material slowly migrating inwards by accretion. The composition of this material is not precisely known and significant variations are found for different IDPs. Hence, for the purpose of the model computations a simplified model composition is introduced which we believe to represent the essential properties of the material. 

\citet{Pol94} have already discussed this problem in the context of calculating opacities for accretion discs. Despite much observational and theoretical efforts to determine the composition of cometary material and large progress in many details \citep[cf. the review of ][]{Mum11} the knowledge on the general composition of the carbonaceous material has not much improved since then. Therefore we base our model essentially on the same assumptions as in  \citet{Pol94}. 

One important parameter is the fraction of the carbon bound in the carbonaceous material. This is estimated from the C/Si ratio in comet Halley particles \citep{Jes89} by \citet{Pol94} to be 0.55 of the total carbon. No better information seems to be available and this estimate does not significantly change if an improved C abundance \citep{Sco09} is used. Therefore we also assume a fraction of 0.55 of the carbon to be condensed in carbonaceous material. The remaining fraction is in the gas phase, and beyond the ice line some of this is frozen.

The carbonaceous material is assumed to be composed of the four components discussed in Sect.~\ref{SectCarbon} carrying the fractions of the condensed carbon given in Table \ref{TabCarbComp} \citep[data taken from][]{Fom97}. With increasing temperature these components decompose, or evaporate, or are destroyed by oxidation, releasing hydrocarbons to the gas phase. The table also shows the typical release temperatures (discussed later) and our assumptions with respect to the gaseous compounds released, which we discuss now.

\subsection{%
Gasification of carbonaceous material}
\label{SectPyrolProc}

An important information for calculating the chemistry is the composition of the gaseous components released on heating the material. We describe the corresponding mixture with an only small number of components because the available information is fragmentary. 

\paragraph{%
Amorphous carbon.}
The oxidation of amorphous carbon is discussed in \citet{Fin97} and \citet{Gai01}. The same reaction model is used here. According to this the carbon is oxidized by an attack of OH radicals (here formed at elevated temperature by thermal dissociation) to carbon atoms at the periphery of aromatically bound  carbon and releasing HCCO molecules to the gas phase. This reaction is irreversible and destroys the carbon.

\paragraph{Aliphatic compounds.}
The aliphatic and aromatic hydrocarbons found in the carbonaceous material are unlikely to exist there as isolated molecules. For the aliphatic compounds the average CH$_2$/CH$_3$ ratio is 2.46 as determined for IDPs  (see Sect.~\ref{SectAnalIDP}) which hints to chain lengths for unbranched chains between 6 and 8 carbon atoms, but the H/C ratio of about 1 found in many cases (see Sect.~\ref{SectAnalCom}) points to highly branched alkanes. They are likely to be part of more complex networks comprising aliphatically and aromatically bound subunits like that shown in \citet{Der10}. 

The outgassing of some laboratory made analogue materials of this kind was studied by \citet{Nak03}. They found that the more volatile part of this material outgasses at temperatures somewhat above 300 K and decomposes during outgassing, releasing short alkanes like CH$_4$ into the gas phase. The release of low molecular weight product gases, dominated by CH$_4$, is also observed for heating of terrestrial kerogen \citep[e.g.][]{Wan13}, though at higher temperatures because of quite different outgassing conditions. Kerogen is often considered as an analogue of the carbonaceous material \citep[cf.][]{Pol94}. The somewhat less volatile aromatically bound component of the material investigated by \citet{Nak03} outgassed at temperatures around 450 K, apparently without decomposition. 

We assume therefore for our model of the carbonaceous material that the aliphatic component is released to some part (50\%) of the corresponding carbon fraction as CH$_4$ (methane), and that the remaining part is released as C$_4$H$_{10}$ (butane) in order to include also a longer alkane chain in our model.  

\paragraph{Aromatic compounds.}
For the aromatic component the weak evidence from observations (see Sect.~\ref{SectAnalIDP}) indicates that released PAHs have up to 16 carbon atoms. We assume in our model for the carbonaceous material that the outgassing of the aromatic component releases C$_{16}$H$_{10}$ (pyrene). This choice is somewhat arbitrary and serves mainly for the purpose to include a small PAH molecule in our model.

Our assumption with respect to the pyrolysis products of the aromatic and aliphatic components are shown in Table~\ref{TabCarbComp}.

\paragraph{CHON material.}
The main component of the carbonaceous material are the CHON particles with very complex composition. They have rather high contents in O, N, and S, e.g., the average composition of the CHON particles in Halley was determined as  C$_{100}$H$_{115}$N$_9$O$_{65}$S \citep{Fom94a} and a composition of C$_{100}$N$_{10}$O$_{50}$ has been given by \citet{Fly08} for the coating of a single grain in an IDP where the composition of the coating could be determined by lucky circumstances. This means that this component carries a non-negligible fraction of the available oxygen (and also of N and others, but we do not consider nitrogen). The material of the CHON particle is supposed to be similar to terrestrial kerogen and, again, we take recourse to what is observed for kerogen pyrolysis. 

Because of the significant oxygen content of the CHON material, CO and CO$_2$ are probably important pyrolysis products of the CHON material. The experimental results of \citet{Nak03} for their proxy of the carbonaceous material shows that most of the oxygen is released in this way. Also some ethanol is released, but this is neglected in our model to keep the number of components low. The nitrogen content of the carbonaceous material is released in the pyrolysis experiments of \citet{Nak03} as NH$_3$. We do not consider any nitrogen and sulphur compounds and subtract from the average composition of the CHON material \citep[C$_{100}$H$_{115}$N$_9$O$_{65}$S,][]{Fom94a} the N and S content and the corresponding number of H atoms and remain with a CHO composition of C:H:O=100:86:65. 

The most abundant carbon bearing highly volatile pyrolysis product of kerogen is CH$_4$ besides less abundant other alkanes and alkenes \citep{Wan13}. Methane is also the most abundant hydrocarbon of low molecular weight in the experiment of  \citet{Nak03}. We assume for simplicity that methane is the sole low-molecular-weight hydrocarbon pyrolysis product. 

Pyrolysis of terrestrial kerogen results to a significant fraction in the production of oil, i.e., in high-molecular-weight compounds. For their proxy of the cometary carbonaceous material, \citet{Nak03} found that some fraction of the pyrolysis products are high-molecular-weight hydrocarbons which did not decompose during evaporation. We assume that the high-molecular-weight component of the pyrolysis products of the carbonaceous material is represented  by the same compound (pyrene) as in the case of the aromatic component of the carbonaceous material in order  to keep the number of components in our model low. 

With the observation that CO$_2$ seems to be somewhat less abundant in the released gases than CO we find the relative fractions for the pyrolysis products  shown in Table~\ref{TabCarbComp} which closely reproduce the assumed approximate C:H:O=100:86:65 ratio for the carbonaceous material.

This is a very simplified model of the carbonaceous material that attempts to describe some basic features of its decomposition into gaseous products at elevated temperature. The further fate of the product molecules has to be determined by modelling the gas phase chemistry of hydrocarbons in the solar nebula. 

\subsection{Pyrolysis temperature}

The laboratory experiment of \citet{Nak03} shows that pyrolysis products are released over some extended temperature regime. The light components are released between about 300 K and 350 K and the heavier products disappear from the residue between about 400 K and 450 K. The data given in the table are confirmed by pyrolysis temperatures derived from kinetic data on the pyrolysis of kerogen as discussed in detail in Appendix~\ref{SectPyrCarb}.
  
This assumption roughly corresponds to the vaporisation temperatures assumed in the dust model of \citet{Pol94} for their volatile and refractory organic compounds, though the reasons for the assumed temperatures are somewhat different. Our assumed pyrolysis temperature are only rather crude estimates. More detailed laboratory investigations are necessary to obtain more precise information on the pyrolysis of the carbonaceous material.

\section{Evolution of the carbonaceous material}


The models of \citet{Gai01} and \citet{Gai02a} already included the oxidation of amorphous carbon dust. In the present model calculation the method for modelling soot oxidation is the same as described in the previous papers and therefore the details are not repeated. 

Besides the amorphous carbon we have to consider additional carbonaceous components, which show a different thermal evolution. According to our model for the carbonaceous material specified by Table~\ref{TabCarbComp}, we have three additional components. It is assumed that these additional components form coatings on the surface of the mineral dust grains (see Sect.~\ref{SectDuStru}) and we assume that these cores all have the same radius $a_{\rm c}$. If the thickness of the coatings is not substantially larger than the core radius, one may neglect the change of surface area during the gasification of the carbonaceous material and assume some average surface area for calculating reaction rates. Then it suffices to consider only the change of the total quantities of the material and not the details of the evolution of the coating.     

Since the gasification of the components releases different gaseous species, it is useful to think of each component as a collection of sub-components with the composition of the corresponding released gas-phase species. For instance, the volatile organic material is assumed to release CH$_4$ and C$_4$H$_{10}$ upon pyrolysis. We introduce two species CH$_4$[v] and C$_4$H$_{10}$[v] that represent the C and H atoms bound in the volatile carbonaceous component that are finally released as CH$_4$ and C$_4$H$_{10}$ molecules to the gas phase. Then the pyrolysis process can be described by the chemical reaction equations
\begin{align*}
\rm CH_4[v]&\longrightarrow\ \rm CH_4 \,,\quad
\rm C_4H_{10}[v]\longrightarrow\ \rm C_4H_{10}
\end{align*}
Analogously we define components C$_{16}$H$_{10}$[m] for the moderately volatile organics and CH$_4$[r], CO[r], CO$_2$[r], and C$_{16}$H$_{10}$[r] for the refractory organic material.

The quantities of each of these components at each location and instant is described by a fictitious concentration $c^\mathrm{(s)}_i$ of this species per H nucleus. The initial concentrations in the low temperature regime of the disk are by definition
\begin{equation}
c^\mathrm{(s)}_{0,i}={f_i\gamma_i\over n_i}\epsilon_\mathrm{C}\,,
\label{InitialConcComp}
\end{equation}
where $\epsilon_\mathrm{C}$ is the element abundance of C with respect to H in the solar nebula, $f_i$ is the fraction of carbon initially bound in the corresponding carbonaceous component to which $i$ belongs, as given in the third column of Table~\ref{TabCarbComp}, and $\gamma_i$ is the fraction  of C atoms released as the particular molecule considered as given in the last column of Table~\ref{TabCarbComp}. Finally, $n_i$ is the number of C atoms in the chemical formula of the species.

The concentrations $c^\mathrm{(s)}_i$ vary with location within the disc and with time by transport and mixing by large-scale flows and turbulent diffusion, and they vary by pyrolysis of the carbonaceous material at elevated temperatures. We therefore add to the set of diffusion-transport-reaction equations described in \citet{Gai01,Gai02a} additional equations for the new components. In cylindrical coordinates and averaged over the vertical direction the equations take the form
\begin{equation}
{\mathrm{d}\,c^\mathrm{(s)}_i\over\mathrm{d}\,t}=
\frac{\partial\,c^\mathrm{(s)}_i}{\partial\,t}+v_{r}\frac{\partial\,c^\mathrm{(s)}_i}{\partial\,r}={1\over hn_\mathrm{H}\,r}\frac{\partial}{\partial\,r}\,r\,hn_\mathrm{H}D\frac{\partial\,c^\mathrm{(s)}_i}{\partial\,r}+{R_i\over n_\mathrm{H}}\,.
\label{TheDiffusionEq}
\end{equation}
Here $v_r$ is the radial drift velocity, $D$ the diffusion coefficient by turbulence, $R_i$ are the reaction rates per unit volume and time, $h$ is the disk height, and $n_\mathrm{H}$ is the number density of H nuclei, for which we have the continuity equation
for the total particle density
\begin{equation}
\frac{\mathrm{d}\,n_\mathrm{H}}{\mathrm{d}\,t}=\frac{\partial\,n_\mathrm{H}}{\partial\,t}+{1\over hr}\frac{\partial\,hrn_\mathrm{H}v_r}{\partial\,r}=0\,.
\end{equation}
In writing down the transport-reactions equations in the form given by Eq.~(\ref{TheDiffusionEq}) we have assumed that mixing processes within the disk by small-scale turbulence-like flows associated with the process driving disk accretion always occurs on much shorter timescales than radial mixing such that the concentrations $c^\mathrm{(s)}_i$ do not vary substantially in vertical direction. This assumption is satisfied as can be seen from Fig.~\ref{FigTauT} where the characteristic timescales
\begin{equation}
\tau_\mathrm{hor}={r^2/D}\,,\quad\tau_\mathrm{vert}={h^2/D}
\end{equation}
for diffusive mixing in horizontal and vertical direction are shown for the inner part of the solar nebula. Additionally it is assumed that chemical processing of the carbonaceous material in the unshielded outermost disk layers by photo-processes is unimportant. \footnote{%
Generally carbonaceous material could be easily processed by photoprocesses as has been shown, e.g., in the work of \citet{Ala14,Ala15}. In protoplanetary disks such processes would be only active in the disk atmosphere, which is not shielded from the ionizing radiation from the protostar, but contains only a small fraction of the total mass. The relative importance of such surface processes in comparison to the processes operating in the disks interior and affecting the overwhelming majority of the material requires 2D disk models including vertical mixing in the chemistry calculation. Such an extension of the model calculation is out of the scope of the present paper.
}%
Such processes are presently not included in our model, but they may form an additional route to carbon oxidation \citep[see][]{Lee10}. Therefore we assume that $c^\mathrm{(s)}_i(r,t)$ represents a vertically averaged concentration of the carbonaceous component~$i$. Additionally it is assumed that there is no substantial relative motion between dust and carrier gas.

The set of Eqs. (\ref{TheDiffusionEq}) has to be solved with appropriate initial and boundary conditions. The initial condition is that we start with concentrations equal to the initial mixture most likely originating from the parent molecular cloud
\begin{equation}
c^\mathrm{(s)}_i(r,0)=c^\mathrm{(s)}_{0,i}\,.
\end{equation} 
The same concentrations are prescribed at the outer disk radius
\begin{equation}
c^\mathrm{(s)}_i(r_\mathrm{ext},t)=c^\mathrm{(s)}_{0,i}\,.
\end{equation} 
The inner radius of the disk model is close to the proto-sun at a distance where all condensed phases are destroyed. Hence we prescribe
\begin{equation}
c^\mathrm{(s)}_i(r_\mathrm{in},t)=0
\end{equation} 
at the inner radius.

The details for evaluating the reaction rates $R_i$ of the pyrolysis processes are described in Appendix~\ref{SectPyr}.

\begin{figure*}

\includegraphics[width=0.32\hsize]{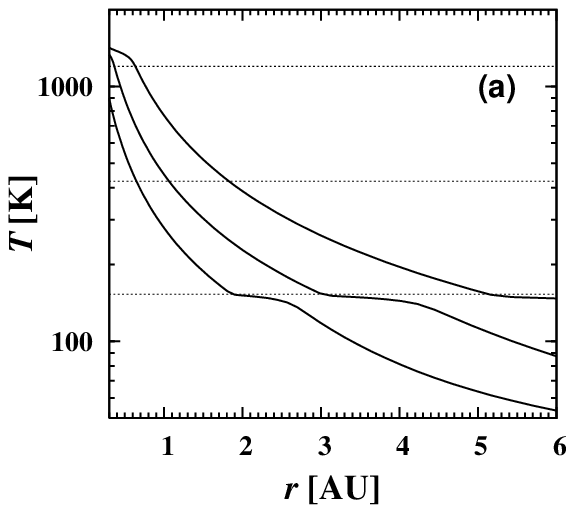}
\hfill
\includegraphics[width=0.32\hsize]{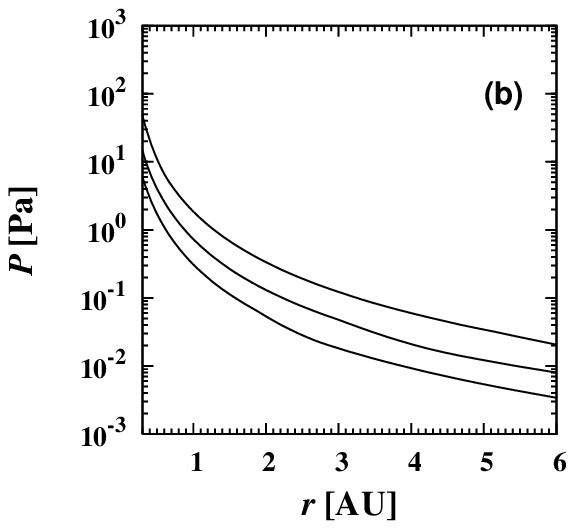}
\hfill
\includegraphics[width=0.32\hsize]{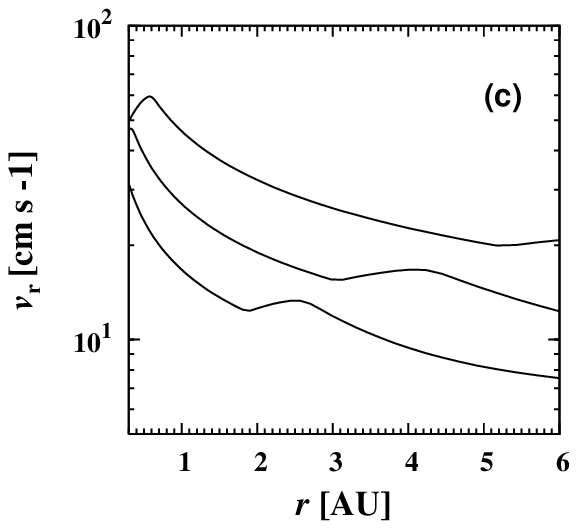}

\caption{Radial variation of some disk properties at three different ages of 1\,Ma, 2\,Ma, and 3\,Ma (from top to bottom). (a) Temperature at the mid-plane. The dotted horizontal lines correspond to the vaporization of ice (152 K), pyrolysis of kerogen (450 K), and oxidation of soot ($\sim1\,200$ K). (b) Pressure at the mid plane. (c) Radial accretion velocity.}

\label{FigDiskProp}

\end{figure*}

\begin{table}

\caption{Parameters of the one-zone disk model}

\begin{tabular}{llll}
\hline
\hline
\noalign{\smallskip}
Quantity & Symbol & Value & Unit \\
\noalign{\smallskip}
\hline
\noalign{\smallskip}
Inner radius & $r_\mathrm{in}$ & 0.1 & AU\\
Outer radius & $r_\mathrm{ext}$ & 200 & AU\\
Stellar mass & $M_*$             & 0.95 & $M_{\sun}$ \\
Disk mass    & $M_\mathrm{disk}$ & 0.05 & $M_{\sun}$ \\
Angular momentum & $J$               &  $1.6\times10^{52}$ & g\,cm$^2$s$^{-1}$             \\
Viscosity parameter & $\alpha$          & $3\times10^{-3}$ &  \\ 
\noalign{\smallskip}
\hline
\end{tabular}

\label{TabParDi}

\end{table}

\section{Carbon abundance in the quiet solar nebula}

The high temperatures required to destroy the carbonaceous component of the pristine disk material are achieved either by moving the material into the hot inner zones of the solar nebula by the accretion flow, or by the flash-like heating events that resulted in chondrule formation. In this section we consider the first process, the gradual heating by accretion; flash heating is considered in the next section.

\subsection{Disk model}
\label{SectDiskMod}

We study the distribution of the carbonaceous material in the quiet solar nebula, i.e., without chondrule formation events, on the basis of a simple one-zone vertically averaged time-dependent model of an accretion disk around a solar-like star. From such models we use the radial variation of temperature and particle density in the disk. This allows to model the chemical processes acting near the midplane of the disk where temperatures and densities are highest in an actively accreting disk and where the chemistry is dominated by neutral-neutral reactions. Processes acting at the disk surface \citep[discussed, e.g., by][]{Lee10} where the material is subject to ionizing radiation driving an ion-molecule reaction chemistry cannot be modeled this way. That requires at least 2D disk models which are out of the scope of the present paper.

\begin{table}[t]

\caption{Element abundances of the solar system, as used in the model calculations (given in the astronomical scale $A(\mathrm{El})=\log\left(n(\mathrm{El})/n(\mathrm{H})\right)-12$).}

\begin{tabular}{lclllll}
\hline
\hline
\noalign{\smallskip}
Element & Abund. & Ref. & & Element & Abund. & Ref. \\
\noalign{\smallskip}
\hline
\noalign{\smallskip}
He & 10.98 & 1 && Mg & 7.53 & 2 \\
C  &  8.47 & 1 && Al & 6.43 & 2 \\
N  &  7.87 & 1 && Si & 7.51 & 2 \\
O  &  8.73 & 1 && Fe & 7.45 & 2 \\
\noalign{\medskip}
$Z$ & 0.142 \\
\noalign{\smallskip}
\hline
\end{tabular}
\tablebib{
 (1) \citet{Asp09}; (2) \citet{Lod09}. 
}

\label{TabSNAbu}
\end{table}

A disk model is calculated for the evolution of the solar nebula. The solution of the model is coupled with a solution of a set of chemical reaction and transport equations, Eq. (\ref{TheDiffusionEq}), which allows to model simultaneously the processing of dust by chemical processes and the mixing of dust material between different disk zones by turbulent flows associated with the viscous process that drives the disk accretion. The model includes the dust components discussed in \citet{Gai04} and the pyrolysis and oxydation of the carbonaceous material. Because the carbonaceous material forms a significant contribution to the opacity of the disk material, the gasification of this component of the condensed matter lowers the opacity and by this back-reacts on the temperature structure of the accretion disk and its evolution. For this reason a consistent coupling between disk evolution model and chemistry is necessary and is included in our model. The details of the model calculation are described in \citet{Gai01} and \citet{Gai02a} and are not repeated here. 
  
The disk model is evolved from some initial state over a time interval of three million years which covers the essential period during which planetesimals are likely to form. The model parameters used for construction of the disk model are shown in Table~\ref{TabParDi}. The models use $\alpha$-viscosity as parameterisation of the process that drives disk accretion. The initial model is constructed such that the region between 0.1 and 30 AU contains a mass $M_\mathrm{disk}$ (the total disk mass is higher, but only the mass in this part of the disk is of relevance for our problem), and an angular momentum $J$ (assumed to be ten times the present angular momentum carried by the present planets). 

The element abundances used for the model calculation are given in Table~\ref{TabSNAbu}. They are required for the opacity calculation which considers the most important mineral components and the carbonaceous material. 

\begin{figure}[t]

\includegraphics[width=\hsize]{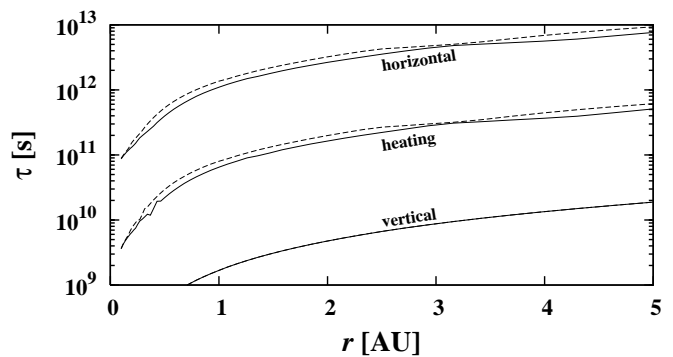}

\caption{Characteristic heating timescales in the inner disk region for a solar nebula model (from Sect.~\ref{SectDiskMod}) at disk ages of 1 Ma (solid line) and 2 Ma (dashed line). Shown are the mixing timescales in horizontal and vertical direction, $\tau_\mathrm{hor}$ and  $\tau_\mathrm{vert}$, respectively, and the timescale for heating of disk material, $\tau_\mathrm{heat}$, due to inward drift by accretion.}

\label{FigTauT}
\end{figure}

Figures \ref{FigDiskProp}a and \ref{FigDiskProp}b show the radial variation of temperature and pressure in the midplane of the disk for the range of distances where the terrestrial planets and bodies from the asteroid belt are located. The results are shown for evolutionary stages of the disk corresponding to ages of 1, 2, and 3 Ma which cover the range of possible formation ages of the bodies in the inner solar system. Figure \ref{FigDiskProp}c shows the accretion velocity which determines how disk matter is transported into regions of increasing temperature. This inward flow of matter rules the thermal metamorphism of the carbonaceous material which we intend to study. The mixing time-scales by turbulent diffusion and the characteristic time-scale for temperature increase by accretion resulting from the disc model are shown in Fig.~\ref{FigTauT}.   

\subsection{Radial variation of carbon abundance}

\begin{figure}

\includegraphics[width=.8\hsize]{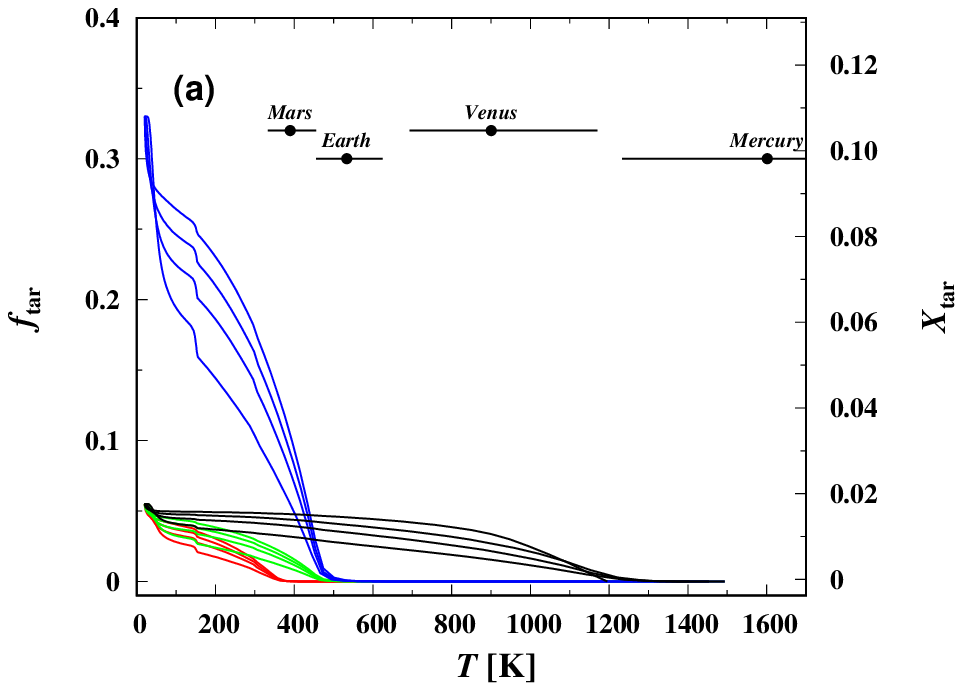}

\includegraphics[width=.8\hsize]{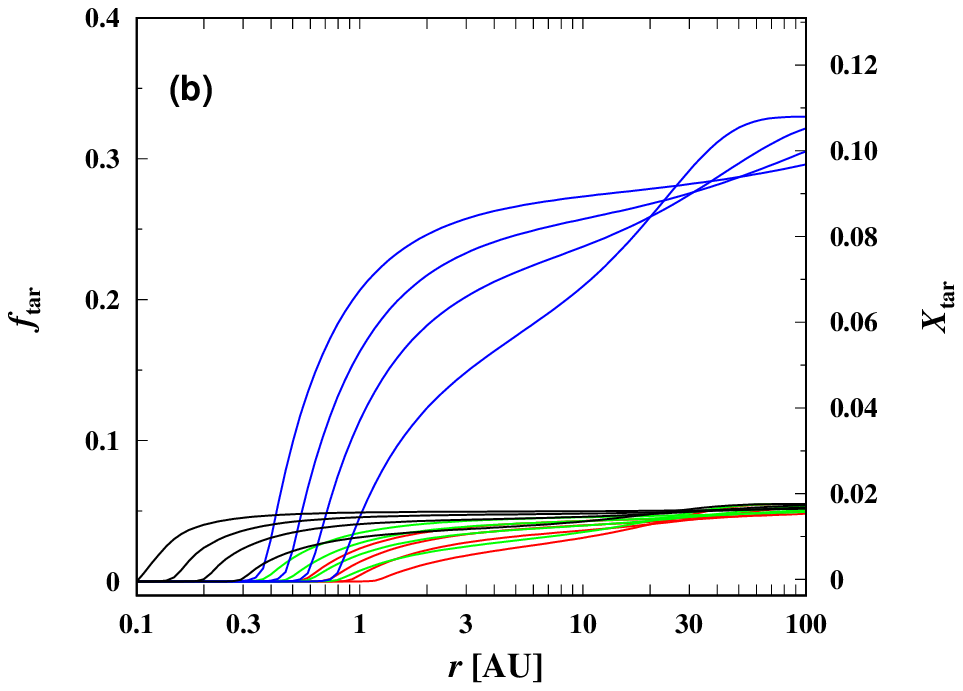}

\caption{(a) Variation with temperature of the fractions (left scale) of the carbon that is bound into one of the solid carbonaceous phase at the midplane of the disk model after 0.2, 0.5, 1 and 2 Ma disk evolution (from lower to upper of the curves in each group, respectively). Black lines: amorphous carbon. Blue lines: refractory organic material (CHON). Green lines: moderately volatile organics. Red lines: volatile organic material. The black dots with horizontal bar indicate the possible range of temperatures in the feeding zones which mainly contribute material to the planets. The right scale shows the mass fraction of carbon contained in dust grains.
(b) Radial variation of the fractions of all components of the carbonaceous material at the midplane of the disk model after 0.2, 0.5, 1, and 2 Ma disk evolution (lower to upper curves for $r<10$ AU).
}

\label{FigFtarT}
\label{FigFtarR}
\end{figure}

Figure \ref{FigFtarT}a shows for $t=0.2$, 0.5, 1, and 2\,Ma after disk formation the result for the abundance variation of the different components of the carbonaceous material (including amorphous soot) with temperature in the mid-plane of the disk. The pyrolysis process depends mainly on the activation energy  and only to a minor extent on the rate of temperature increase [see Eq.~(\ref{Tpyr})] as the matter spirals-in into regions of higher temperature during disk accretion. For this reason the variation over time of the fractional abundance curves with temperature for each of the materials are moderate to small. It is easy to recognize how the different materials disappear at their characteristic pyrolysis temperatures. 

The dots with error bars show the temperature in the zone from which the planets acquire most of their material according to the model of \citet{Lew74}. This considers the varying composition of solids in equilibrium with each other and the gas phase with varying temperature (and pressure) and compares this with the composition of the planets. Though the assumptions on which this model resides are now strongly questioned, it may still serve as a hint to the relevant temperature range of in the solar nebula from which the planets assembled the bigger part of their  material. The model calculation shows that in the main feeding zones of the terrestrial planets the volatile carbonaceous components are already lost to the gas phase by pyrolysis, but that the amorphous carbon component fully contributes the material that forms the planetesimals in the formation zone of Mars to Venus. Only the formation zone of Mercury would be carbon free due to oxidation. 

Figure \ref{FigFtarR}b shows for the same four evolutionary periods the variation of the fractional abundances with radius. In this representation one sees clear evolutionary effects because the disk looses matter to the star and becomes gradually cooler as the surface density decreases. This shifts all the abundance curves inward which remain, however, essentially self-similar such that in the representation in Fig.~\ref{FigFtarT}a the different disk evolutionary stages appear almost identical. 

From these figures one infers that the resulting carbon content of the planetesimals in the formation zone of Mars to Venus would correspond to a mass fraction of serveral percent. This strongly exceeds the carbon content estimated in models for the mantle composition of terrestrial planets. Further, the carbon content of, e.g., ordinary chondrites which are assumed to originate from the region $\lesssim2.5$ AU would exceed 10\%, which is also in conflict with observations. The low carbon content of the terrestrial planets and ordinary chondrites, hence, is not explicable by gradual warming and associated decomposition and chemical reduction to CO during accretion flow and requires a different process.

\section{Carbon destruction by flash-heating}

Except for the CI chondrites all chondrite classes contain chondrules which are thought to be the result of flash-like heating processes acting in the solar nebula. Despite considerable effort in cosmochemistry the process or processes responsible for their formation remains unclear, but these investigations provide a lot of information on the thermal history of chondrules derived from the findings with respect to their composition and texture. These investigations showed \citep[see][ for reviews]{Jon00,Cie05} that they were heated to high temperatures with peak values in the range between about 1\,600 and 2\,100 K. The main facts are: (1) The chondrules are heated very rapidly and stay for only a brief period, a few minutes, at their peak temperature. (2) They start heating from low temperatures. (3) They cool down after the flash-like heating at rates of 10 K/hr to 1\,000 K/hr. (4) Chondrule formation events occur repeatedly. 

\begin{figure}

\includegraphics[width=\hsize]{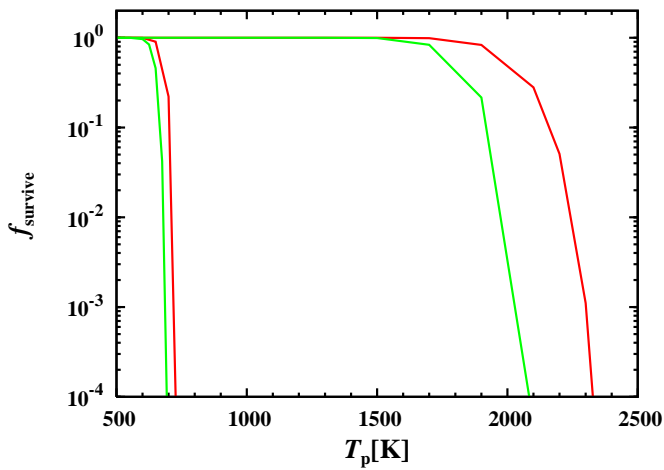}

\caption{Fraction of refractory anthracene (left) and amorphous carbon (right) that survives flash-heating during chondrule formation for different cooling times: $\lambda^{-1}=10$ h (red lines), $\lambda^{-1}=100$ h (green lines).}

\label{FigFlashSur}

\end{figure}

\begin{figure*}
\sidecaption
\includegraphics[width=12cm]{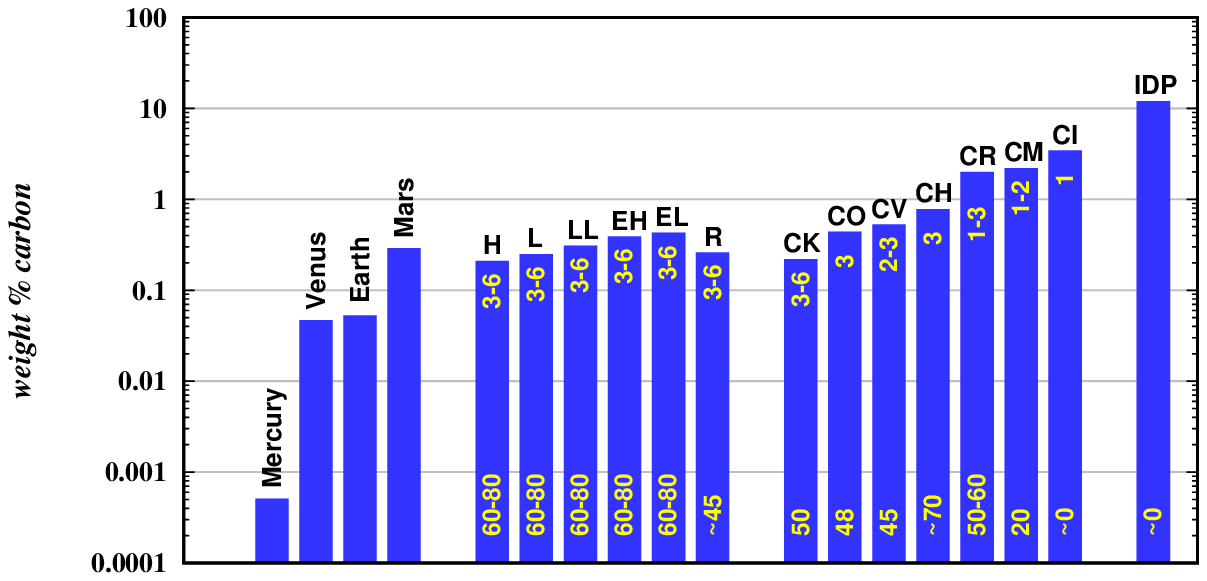}
\caption{Carbon content of bodies in the terrestrial planet region, the Asteroid Belt, and in cometary particles (anhydrous porous interplanetary dust particles). The lower numbers in the bars give the percentage of volume filled by chondrules, the upper numbers the range of petrologic types. For sources of data see text.}

\label{FigCarXfrac}
\end{figure*}

It is to be expected that the temperature evolution reconstructed for the chondrules also reflects that of the ambient gas. The carbonaceous component of any dust aggregate initially present in the disk would start to decompose and oxidize (Appendix~\ref{SectGasChem}) once it experiences a flash-heating event. The important question then is whether the duration of the high-temperature spikes responsible for chondrule formation is sufficiently long that carbon is completely destroyed or does part of it survive such events. To check this we solve the system of rate Eqs. (\ref{TheDiffusionEq}) without mixing and transport terms (these are negligible over the short duration of a flash event) for a fixed mass density corresponding to a pressure of $10^{-4}$\,bar at $300$\,K. This pressure is typical for the region around 2\,AU in the solar nebula (see Fig.~\ref{FigDiskProp}b) where the chondrite parent bodies formed. The temperature is varied according to the following prescription:

\begin{enumerate}

\item The temperature is increased linearly within 1 minute from 300\,K to some maximum temperature $T_\mathrm{p}$.

\item The maximum (plateau) temperature  $T_\mathrm{p}$ is held fixed for 10 minutes.

\item The temperature decreases exponentially from $T_\mathrm{p}$ with decay constant $\lambda$ of several hours.

\end{enumerate}
This is taken as an approximation to the temperature spikes that are  responsible for chondrule formation. The plateau temperature  $T_\mathrm{p}$ is varied between 1\,600\,K and 2\,100\,K as is suggested by the results of investigations on chondrules. The initial conditions for the gas phase and the carbonaceous components are as in Appendix~\ref{SectGasChem}. The reaction network is supplemented in this case by the reaction
\begin{displaymath}
\rm C_2[s]\ +\ O\ \longrightarrow\ C_2O
\end{displaymath}
because contrary to the quiet disk the peak temperature during chondrule-forming flash-heating is sufficiently high that O atoms may become abundant gas phase species that contribute to carbon oxidation.

Figure \ref{FigFlashSur} shows the fractions of the initial content that survive a single flash event for two of the  carbonaceous components, amorphous carbon and anthracene from the refractory CHON component, for varying plateau temperatures and different decay times $\lambda^{-1}$ after passing the peak temperature. The amorphous carbon is rather resistant. Only flash-heating events with the highest peak temperatures destroy the amorphous carbon. The next refractory of the carbonaceous components, the refractory anthracene from the CHON material, would only survive heating events with temperatures less than 700 K. Hence, this and all other less refractory components of the pristine carbonaceous material are completely destroyed by the flash-heating events occurring in the solar nebula as documented by the existence and properties of chondrules.

\section{The carbon content of planetesimals}

From our calculations it follows that in the quiet disk the pristine carbonaceous component of the dust material that is transported inwards during the accretion process is gasified in the warm inner parts of the solar nebula. The volatile and refractory hydrocarbons disappear between 300 K and 500 K by pyrolysis while the minor component, the amorphous carbon, survives up to about 1\,200\,K until this also disappears by oxidation with water vapour. Once the planetesimal formation process commences in the solar nebula the planetesimals would initially be endowed with significant amounts of solid carbonaceous material between several mass percents in the inner terrestrial planet region and up to more than about 10 mass percent in the asteroidal region. This is much more than what is actually found in their present descendants.  

The second important process that results in efficient destruction of carbonaceous material is the process that results in the formation of chondrules. The hydrocarbon components of the carbonaceous components of the pristine dust material are all gasified during flash-heating, except for the amorphous carbon. The amorphous carbon is only partially destroyed in a single such an event and this happens only for the flash-heating events that reach the highest peak temperatures that are inferred from chondrule properties. Investigations of chondrules revealed that the flash heating is a repeated process because many chondrules show indications that they contain material from former chondrule forming episodes \citep[e.g.][]{Jon00}. In those regions of the solar nebula where this process was active the carbon content of the dust will be strongly reduced to values below the small initial fraction of amorphous carbon or the dust material is even more or less completely cleared from carbon.  

In the case of ordinary chondrites, about 80\% of their parent bodies were composed of chondrules and contained only a small residual fraction of dust (= matrix). The disk material in this region unavoidably must have been subject of many repeated flash heating events, which means that the initial carbon
content must have been nearly completely destroyed. But this does not mean  that this region is completely cleared from carbon, because some matrix material may have escaped flash heating events, and/or the extended chondrule formation processes may have extended over hundreds of thousends of years which leaves enough time to mix some fresh carbon bearing material into this region from outside. Hence, one expects in this region a small but non-zero carbon content of the disk material at the instant when the solid material is assembled into planetesimals.

A quantitative statement on the resulting carbon content can presently not be made. A model calculation referring to this requires to know the frequency and exact intensity of the flash-heating events which are presently not known.  
 
A different situation is obviously encountered in the disk region beyond the ice-line where the carbonaceous chondrites formed. The CI chondrites are practically chondrule free and the CM chondrites contain only a low fraction of chondrules. This means that the process responsible for the flash-heating events that formed chondrules did not work in this region or it is rather inefficient such that flashes are rare, but still individual flashes are sufficiently energetic to produce typical chondrule textures.

That chondrule flash heating events and chondrule abundances control the carbon content of chondrites and their parent bodies is immediately supported by the observation that chondrules are highly depleted in carbon, while the matrix component is the carbon carrier, particularly in carbonaceous chondrites \citep[e.g., ][]{Mak93, Ale07, Pea06}.

Such general trends are seen in the different meteorite classes. Figure \ref{FigCarXfrac} shows the carbon content of the different chondrite classes and that of anhydrous IDPs which are derived from comets. For the different chondrite classes the volume fraction of chondrules is also noted.  the compilation of \citet[ Table 1, based on the sources cited therein]{Hen16} for the chondrites and from \citet{Mor79,Mor80}, \citet{Mar12}, and \citet{McD94} for the planets. The carbon abundances are from \citet{Jar90} and \citet{Pea06}. The chondrule abundance of CK chondrites is given in the literature as 15\%, but a new determination by \citet{Cha16} found $\sim50$ volume percent for CK3 meteorites. It has been proposed that CK and CV are derived from the same parent body \citep{Gre10} and this may be the parent body of the Eos family \citep{Gre10a,Nov14}. The chondrule abundance for CH chondrites is taken from \citet{Hez92}.

Obviously there is some general trend that the chondrule abundance decreases with increasing carbon content which fits to the expectations following from the carbon gasification and flash heating processes acting in the solar nebula. 
Specificly, the following groups can be distinguished:
\begin{enumerate}

\item IDPs. Their carbon content represents the initial state which is found at distances $\gtrsim10$\,AU where the comets formed. Chondrule fragments occasionally found in IDPs are thought to be formed in the warm parts of the solar nebula and transported outwards \citep[e.g.][]{Bro14}.

\item The CI and CM chondrites, where chondrules have low abundance or are almost absent. Their high water content indicates that they are formed in the trans-snowline region of the solar nebula. Their carbon abundance is lower than in IDPs, which can partly be ascribed to their formation at different disc radii and parent body related processes.

\item The CV + CK and CO chondrites and also the R chondrites, where chondrules are abundant but not dominant and for which the carbon content is significantly lower. Their low water content indicates that they were formed not too far away from but inside of the snowline region of the solar nebula. 

\item The ordinary chondrites, with very abundant chondrules, absence of water, and a somewhat smaller carbon abundance than the previous group. They are obviously formed in a region where repeated flash-heating events destroy the carbon carried inwards by the accretion process.

\item The terrestrial planets with a significantly lower carbon content than chondrites. They form in a region where only the small fraction of amorphous carbon could survive in the quiescent nebula, but almost most of this is likely to be destroyed already by flash heating since mass-accretion in the disk requires the material first to pass through the region $r\gtrsim2$\,AU before it enters the terrestrial planet forming region. The small surviving fraction of the carbon would be included in the planetesimals and the planets forming from them. During the differentiation into core and mantle this carbon could partly be dissolved in the core material, leaving a mantle depleted in carbon. This is derived from recent studies that the outer core of the earth may contain some carbon at the 1\% weight level \citep{Nak15,Woo13}.

\item Mercury which formed in a region of the solar nebula where also the resilient amorphous carbon is destroyed by oxidation such that no condensed carbon exists in this region.

\end{enumerate}

However, there are also some obvious discrepancies hinting at other or additional processes governing the carbon content: 

\begin{enumerate}

\item  The CR and CH chondrites with their high chondrule content do not easily fit into this picture. They belong to a common meteorite clan formed by CR, CH, and CB  \citep{Kro14}. The CB chondrites are not considered here because they are probably formed as the product of an asteroidal collision \citep{Fed15}. The same seems also to hold for CHs \citep{Kro10} such that their properties are not solely determined by Solar Nebula processes.

\item Loss of carbon on parent bodies due to high temperature metamorphism. For example, \citet{Ale07} found that in thermally metamorphosed type 3 ordinary and CV chondrites carbon could be depleted by a factor of 2-3, similar to earlier work by \citet{Mak93}. Similar, differentiation, possibly affecting the precursor planetesimals of terrestrial planets, could also have caused significant carbon loss. 

\item  Reduced assemblages have generally higher carbon contents, e.g., enstatite chondrites when compared to ordinary chondrites, or CH when compared to CO chondrites. 

\end{enumerate}

\section{Concluding remarks}
\label{SectConclu}

We modelled the processing of carbonaceous matter in the solar protoplanetary disc. Much of the carbon originally contained in solid phases is gasified and lost and cannot be incorporated into planetary bodies, particularly those forming in the inner disc. However, the carbon depletion of ordinary chondrites and terrestrial planets is higher than expected from combustion at typically modelled disc temperatures. Chondrule flash heating is an effective additional process causing carbon depletion in chondritic planetesimals, a result which is corroborated by the fundamental observation that the main carbon carrier in chondritic meteorites is the fine grained matrix, while chondrules are generally poor in carbon.

While it is not completely ascertained inhowfar the precursor material of the terrestrial planets contained chondrules, the ubiquity of chondrules in undifferentiated planetesimals  indicate that chondrule forming processes were active throughout the solar nebula, particularly in the inner disc. While the incorporation of carbon poor chondrules into precursor planetesimals is an essential factor of carbon depletion, further reduction of the carbon content may have been due to thermal processing on planetesimals due to heating by $^{26}$Al decay heat, which lead to strong thermal metamorphism or even differentiation into metallic cores and silicate mantles of even relatively small (about 100 km sized) planetesimals.


\begin{acknowledgements}
This work was supported by `Schwerpunktprogramm 1385' supported by the `Deutsche Forschungs\-gemeinschaft (DFG)'. MT acknowledges support by the Klaus Tschira Stiftung gGmbH. This research has made use of NASA's Astrophysics Data System.

\end{acknowledgements}

\begin{appendix}

\section{Rate coefficients for pyrolysis}
\label{SectPyr}

\begin{table}

\caption{Data for pyrolysis of kerogen \citep[from][]{Oba02}.}

\begin{tabular}{llcll}
\hline
\hline
\noalign{\smallskip}
Released species & \multicolumn{1}{c}{$A$} & \multicolumn{1}{c}{$E_\mathrm{a}$} & \multicolumn{1}{c}{$E_\mathrm{a}/R$} & \multicolumn{1}{c}{$T_\mathrm{py}$} \\
\noalign{\smallskip}
                & \multicolumn{1}{c}{[s$^{-1}]$} & \multicolumn{1}{c}{[kJ/mol]} & \multicolumn{1}{c}{[K]} & \multicolumn{1}{c}{[K]} \\
\noalign{\smallskip}
\hline
\noalign{\smallskip}
Benzene & $2.46\times10^{12}$ & 205.0 & 24\,650 & 450 \\
Hexane & $2.60\times10^{14}$ &  223.0 & 26\,810 & 450 \\
Indene & $4.90\times10^{11}$ &  195.4 & 23\,500 & 440 \\
\noalign{\smallskip}
\hline
\end{tabular}

\label{TabPyrol}
\end{table}

\subsection{Pyrolysis of carbonaceous material}
\label{SectPyrCarb}

Next we have to specify the reaction rates $R_i$ for the pyrolysis processes considered in Sect.~\ref{SectPyrolProc}. If we describe the species bound in one of the components of the carbonaceous material by a fictitious concentration $c^\mathrm{(s)}_i$ and if it is assumed that the decomposition process can be described as a homogeneous first order reaction process then the reaction rate can be written as
\begin{displaymath}
R_i=k_ic^\mathrm{(s)}_in_\mathrm{H}
\end{displaymath}
with reaction rate coefficient $k_i$ in units per unit time.

We estimate the rate coefficient $k_i^{(\mathrm{c})}$ from results of laboratory experiments on pyrolysis of kerogen. The observed rates have been shown to be interpreted best of all by assuming that (1) the pyrolysis process is a homogeneous first order rate process and (2) that $k_i$ is described by a series of Arrhenius-law like terms
\begin{equation}
k_i=A_i\,{\rm e}^{-{E_{i,\mathrm{a}}/RT}}
\label{RatPyrArr}
\end{equation}
with activation energies $E_{i,\mathrm{a}}$ from a certain range of values \citep{Bur87}. Because of the expected structural relationship of the carbonaceous material with kerogen one can speculate that also for the carbonaceous material of IDPs one needs such a description to correctly account for the liberation of pyrolysis products over an extended temperature range, which is also seen in the experiments of \citet{Nak03}. On the other hand, presently only insufficient information is available to construct such refined kind of models. Therefore we approximate the rate $k_i$ by a single law of the form of Eq.~(\ref{RatPyrArr}).

\begin{table}

\caption{Kinetic data assumed for the pyrolysis of the carbonaceous material in the solar nebula.}

\begin{tabular}{lccc}
\hline
\hline
\noalign{\smallskip}
Component & $A$ & $T_0$ & b \\
\noalign{\smallskip}
                & [s$^{-1}]$ & [K] & \\
\noalign{\smallskip}
\hline
\noalign{\smallskip}
Volatile organic material & $4.00\times10^{13}$ & 19\,050  \\
Moderately volatile organics & $4.00\times10^{13}$ &  24\,500 \\
Refractory organic material & $4.00\times10^{13}$ & 24\,500  \\[.2cm]
CH$_4$[v] &  $2.00\times10^{13}$ & 19\,050  \\
C$_4$H$_{10}$[v] & $5.00\times10^{12}$ & 19\,050 \\
C$_{16}$H$_{10}$[m]  & $2.50\times10^{12}$ & 24\,500 \\
CH$_4$[r] &  $6.00\times10^{12}$ & 24\,500 \\
CO[r] &  $1.00\times10^{13}$ & 24\,500 \\
CO$_2$[r] &  $8.00\times10^{12}$ & 24\,500 \\
C$_{16}$H$_{10}$[r]  & $1.00\times10^{12}$ & 24\,500 \\[.2cm]
C$_2$[s] & $1.00\times10^{-15}$ & & 0.5 \\
\noalign{\smallskip}
\hline
\end{tabular}

\label{TabKinCarbComp}
\end{table}

In a study on kerogen pyrolysis \citet{Oba02} obtained values for the prefactor $A$ and activation energy $E_\mathrm{a}$ for some important pyrolysis products. These are given in Table~\ref{TabPyrol}. Less specified data given in \citet{Bur87} are similar. We calculate as in \citet{Chy90} a release temperature from the relation
\begin{equation}
T_\mathrm{py}={E_\mathrm{a}\over R\ln(\tau_\mathrm{heat}A)}\,,
\label{Tpyr}
\end{equation}
where $R$ is the gas constant and $\tau_\mathrm{heat}$ a characteristic heating time. This is the temperature where from the initial material the fraction $1/e$ is released by pyrolysis if the material is held at this temperature for the period $\tau_\mathrm{heat}$. Figure \ref{FigTauT} shows for a model of the solar nebula the timescale for a substantial temperature change defined here to be
\begin{displaymath}
\tau_\mathrm{heat}=0.1\,r/v_\mathrm{r}
\end{displaymath}
with $v_r$ being the inward drift velocity in an accretion disk. A typical value of $\tau_\mathrm{heat}$ is $1\dots3\times10^{11}$s at distances in the range 1\dots3 AU. Typical release temperatures  $T_\mathrm{py}$ for some pyrolysis products calculated from this are shown in Table~\ref{TabPyrol}.

The calculated release temperatures are comparable to the findings of laboratory experiments of \citet{Nak03} for their proxy for the cometary carbonaceous material, though the heating periods in the experiments where shorter, which means that the bond energies in the laboratory-made material are somewhat lower than for natural kerogen. It is to some extent an accidental coincidence that despite of different conditions in the experiment and for disk models the temperatures turn out to be about the same.

We will use the kerogen-based data in our model calculation for the pyrolysis of the CHON material. The rate term in Eq.~(\ref{TheDiffusionEq}) is 
\begin{equation}
{R_i\over n_\mathrm{H}}=A_i\,{\rm e}^{-T_{0i}/T}\,c^\mathrm{(s)}_i
\end{equation}
for each of the components of the carbonaceous material (except soot) from Table~\ref{TabCarbComp}. A single value for $A_i$ and $T_{0i}$ is assumed for simplicity.  The values of $T_{0i}$ are slightly adjusted such that the pyrolysis temperatures given in Table~\ref{TabCarbComp} are reproduced. The rate coefficients for the loss of C atoms by pyrolysis for three components of the carbonaceous material are shown in the first three rows of Table~\ref{TabKinCarbComp}. The following row gives the coefficients for the pyrolysis rates of the individual products. They consider the relative fraction of carbon released by the different species, as given in Table~\ref{TabCarbComp}, and the number of C atoms carried by them. It would be possible to account in more detail for the data of individual pyrolysis species given by \citet{Oba02}, but in view of our only fragmentary present knowledge of the astronomical pyrolysis process we feel it presently not justified to consider more details.  
 
\begin{figure*}[t]

\centerline{
\includegraphics[width=.40\hsize]{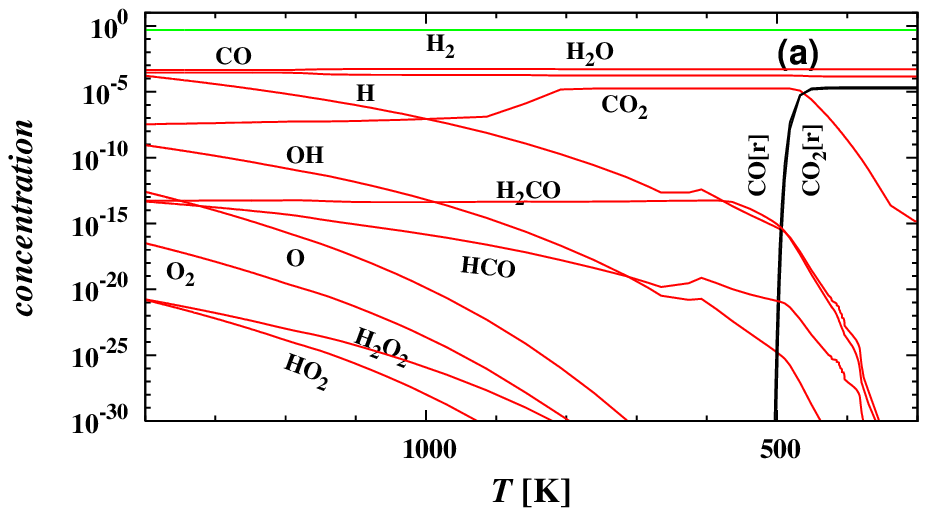}
\ 
\includegraphics[width=.40\hsize]{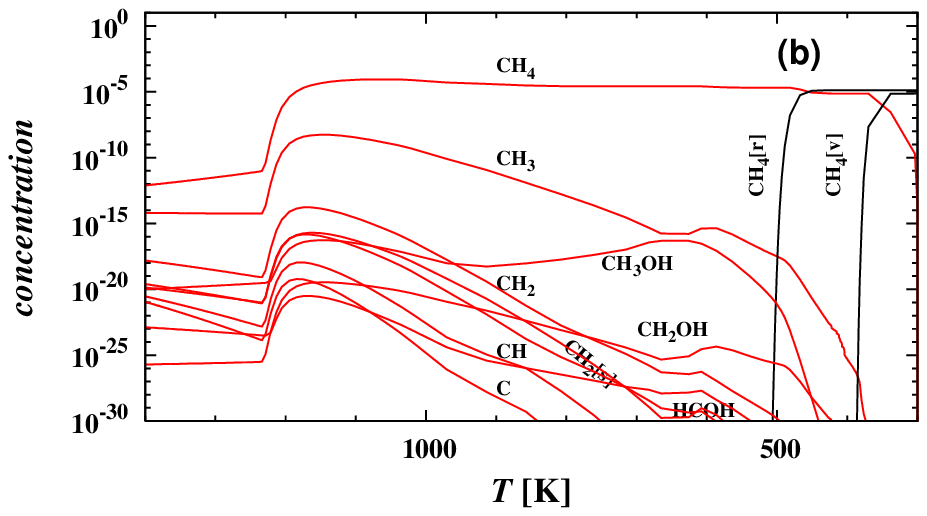}
}

\centerline{
\includegraphics[width=.40\hsize]{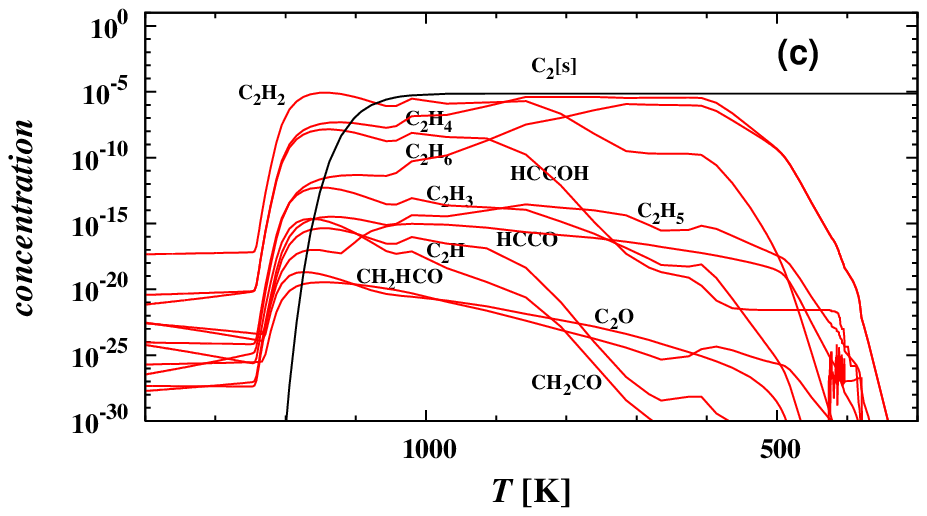}
\ 
\includegraphics[width=.40\hsize]{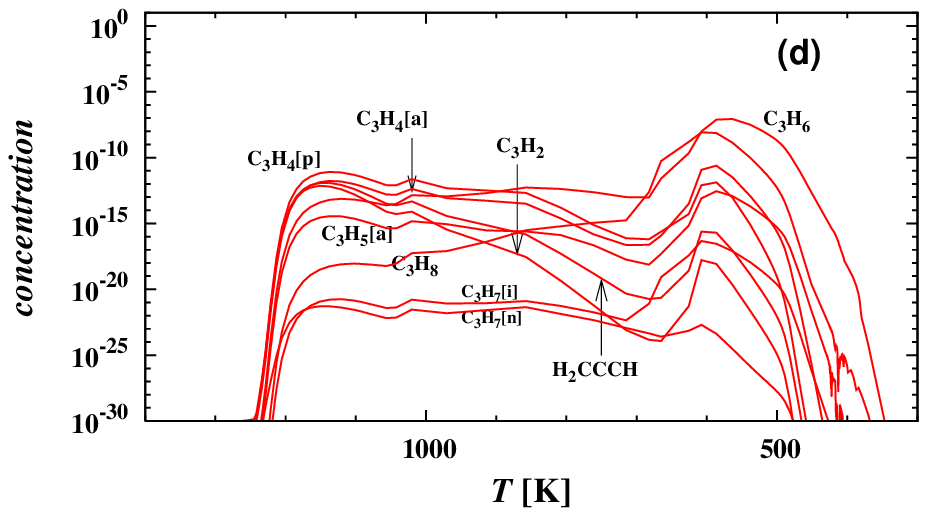}
}

\centerline{
\includegraphics[width=.40\hsize]{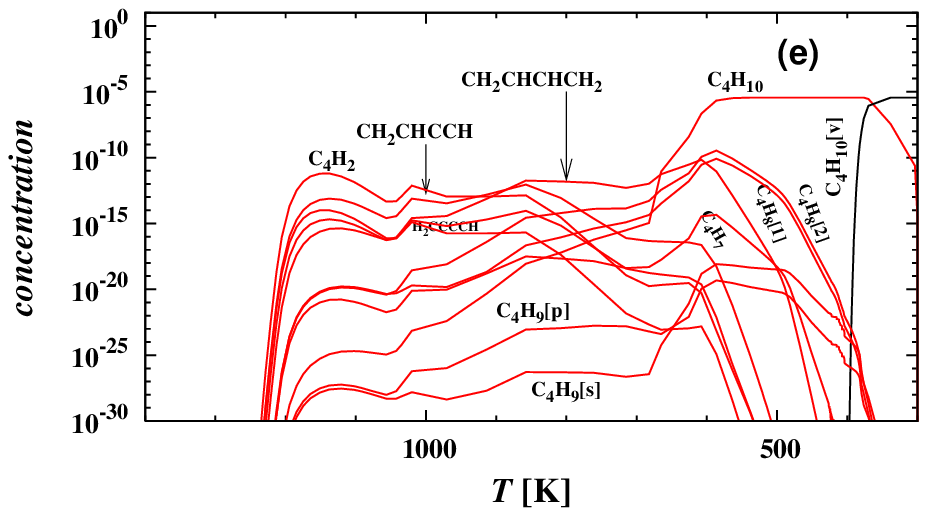}
\ 
\includegraphics[width=.40\hsize]{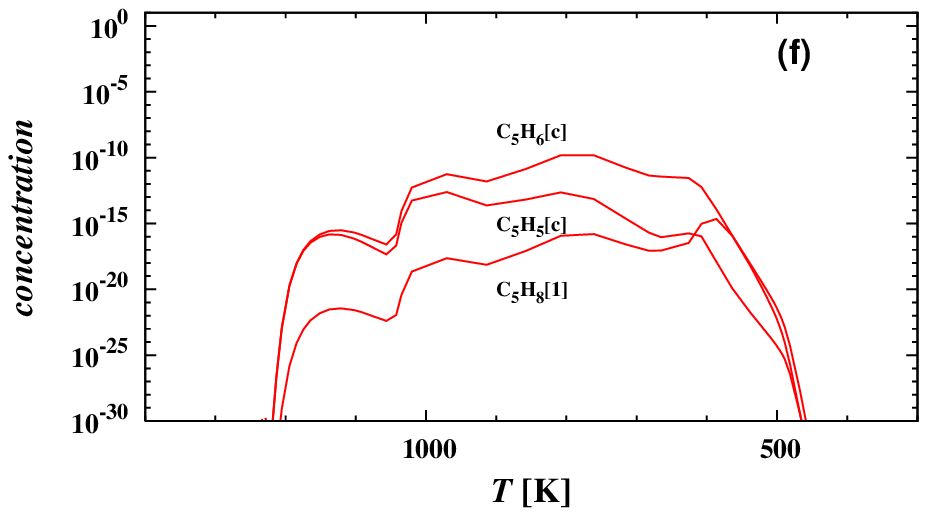}
}

\centerline{
\includegraphics[width=.40\hsize]{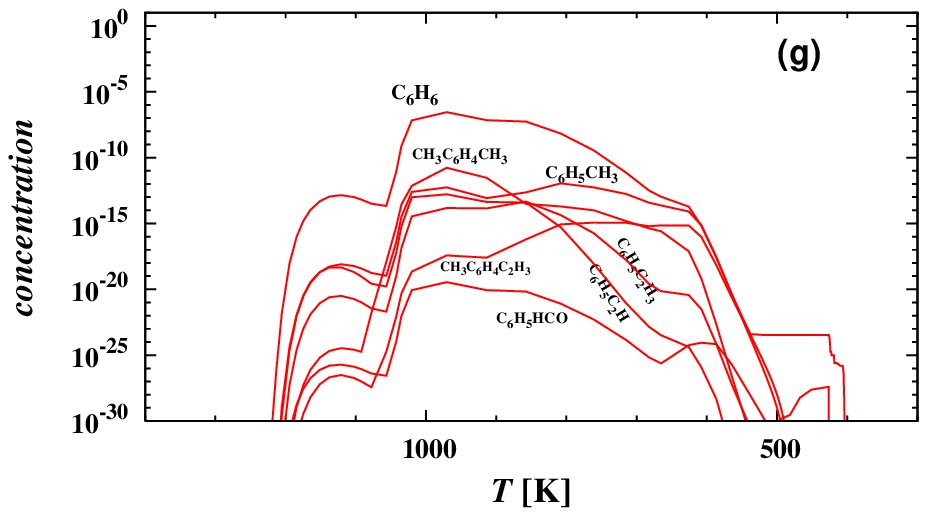}
\ 
\includegraphics[width=.40\hsize]{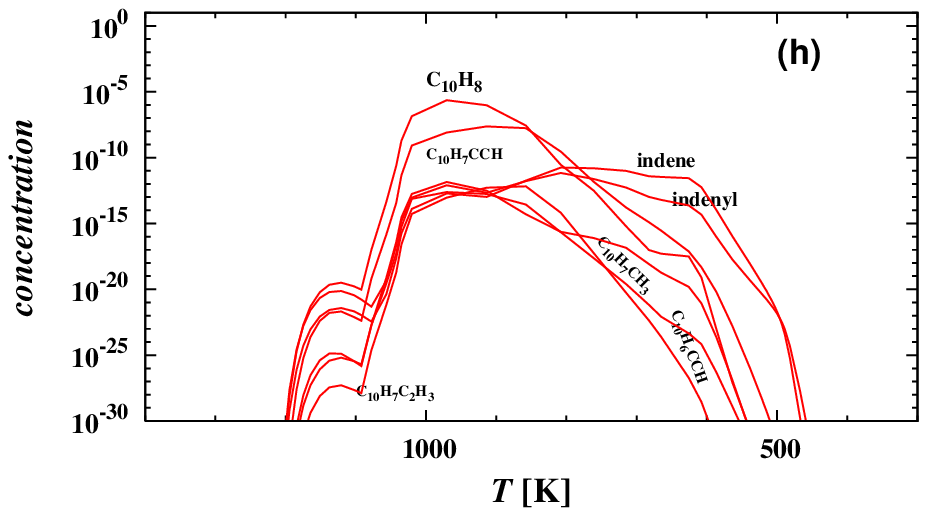}
}

\centerline{
\includegraphics[width=.40\hsize]{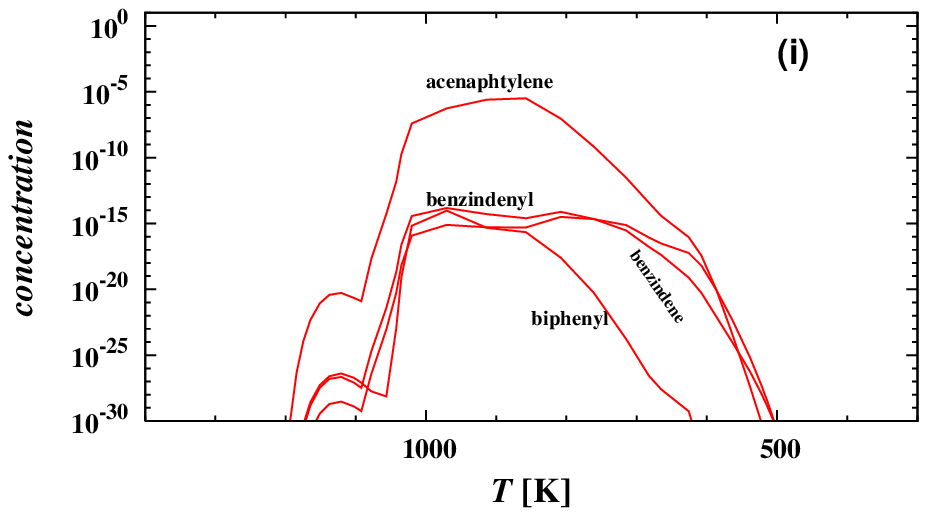}
\ 
\includegraphics[width=.40\hsize]{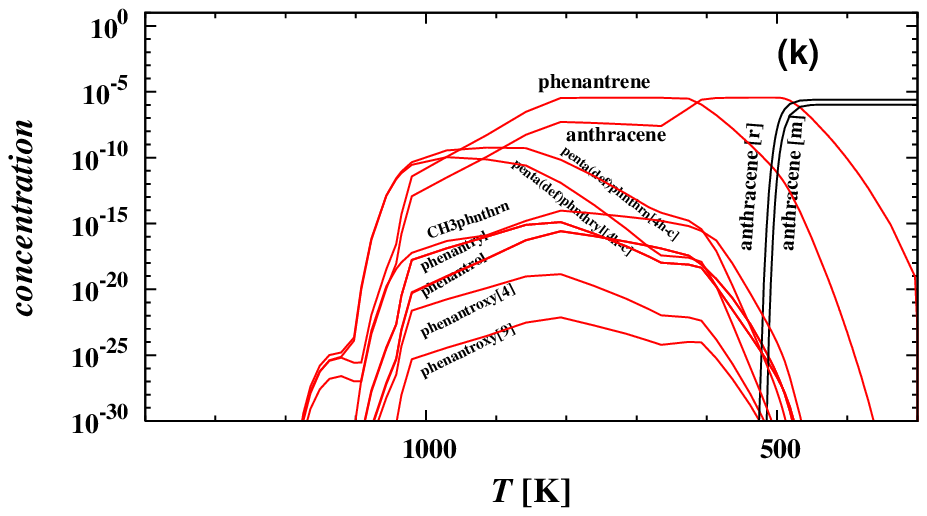}
}

\caption{Concentration relative to H nuclei of the most abundant species of the HCO chemistry for oxidation of the carbonaceous material in the solar nebula to CO by water vapour. Note that the temperature increases from the right to the left. (a) The water chemistry and the chemistry of CO$_2$ and H$_2$CO. (b) -- (f) The chemistry of  hydrocarbons with one to five C atoms. (g) The chemistry of hydrocarbons with six and seven C atoms.  (h) The chemistry of hydrocarbons with 8 to 10  C atoms.  (i) The chemistry of hydrocarbons with 12 and 13  C atoms. (k) The chemistry of hydrocarbons with 14 and 15 C atoms. The black lines correspond to the components of the solid carbonaceous material. }

\label{FigPyrolCarbMat} 
\end{figure*}

\subsection{Oxidation of soot}

The oxidation of carbon particles is a process which has extensively been studied in flame chemistry. In  \citet{Dus96} and \citet{Fin97} the results for the basic oxidation mechanism of solid carbon obtained in the chemistry of flames have already been applied to the problem of destruction of carbon dust grains in protoplanetary accretion discs. Under conditions encountered in the early solar nebula the basic reaction scheme for oxidation of solid carbon into CO starts with the reaction where a OH radical attacks a six-ring at the periphery of a large PAH and cracks carbon-carbon bonds. The reaction is assumed to be of the type
\begin{displaymath}
\rm soot\ +\ OH\ \longrightarrow\ soot-2C\ +\ HCCO\,.
\end{displaymath}
This can be cast into the form of a chemical reaction equation if we introduce a fictitious species C$_2$[s] bound in solid carbon material. The oxidation reaction can be written with this assumption as
\begin{equation}
\rm C_2[s]\ +\ OH\ \longrightarrow\ HCCO\,.
\label{ReactSootOxi}
\end{equation}
If the carbon dust particles are spheres of radius $a$, the relation between the particle density of the species C$_2$[s] and the number density $n_\mathrm{car}$ of carbon grains is
\begin{equation}
{4\pi a^3\over 3}\rho_\mathrm{car}\,n_\mathrm{car}={2A_\mathrm{C}m_{\rm H}}\,n_\mathrm{C_2[s]}^\mathrm{(s)}\,,
\end{equation}
where $A_\mathrm{C}$ is the atomic weight of carbon and $\rho_\mathrm{car}$ the mass density of solid carbon.

The rate for the starter reaction of soot oxidation is
\begin{equation}
{R_\mathrm{car}\over n_\mathrm{H}}=k_\mathrm{car}n_\mathrm{OH}c^\mathrm{(s)}_\mathrm{C_2[s]}
\end{equation}
with $c^\mathrm{(s)}_\mathrm{C_2[s]}$ being the particle density of C atoms bound in soot, $n_{\rm OH}$ the particle density of OH radicals, and $k_\mathrm{car}$ the corresponding rate coefficient which can be written as
\begin{equation}
k_\mathrm{car}={3v_\mathrm{OH}\alpha2A_\mathrm{C}m_\mathrm{H}\over a\rho_\mathrm{car}}\,.
\end{equation}
Here $v_{\rm th,OH}$ is the r.m.s. thermal velocity of OH,
\begin{equation}
v_{\rm th,OH}=\sqrt{kT\over2\pi m_{\rm OH}}\,,
\end{equation}
and $\alpha_{\rm car}$ (= 0.1) is the efficiency for the reaction of OH with solid carbon \citep[see][for this]{Fin97}. Writing this as
\begin{equation}
k_\mathrm{car}=AT^b
\end{equation}
we obtain with an assumed radius of $a=5\times10^{-6}$ (see Sect.~\ref{SectAnalCom}) and density $\rho_\mathrm{car}=2\,\rm g\,cm^{-3}$ the numerical coefficients $A$, $b$ for the rate coefficient given in Table~\ref{TabKinCarbComp}.

\section{Gas phase chemistry of hydrocarbons}

The oxidation of hydrocarbons is the basic process responsible for the final conversion of the pyrolysis products into CO. Because of their technical importance the burning processes of organic materials are well studied within in the field of flame chemistry \citep[e.g.][]{War06}. The main difference between terrestrial flames and processes in an accretion disc is the high excess of hydrogen in astrophysical settings. Therefore most oxygen not bound in CO is found in H$_2$O and its dissociation products OH and O and almost no O$_2$ is found, which is a major oxygen carrier in technical flames. Also CO$_2$ is of low abundance contrary to terrestrial flames. Also the pressure is lower in the solar nebula than in laboratory low-pressure flames, but not by orders of magnitude, and temperatures at which oxidation occurs are expected to be lower because of much longer characteristic timescales for temperature changes. The general conditions, however, under which the chemistry operates are similar.  This means that the relative importance of the various oxidation reactions are different in the solar nebula from that in the laboratory, but the same suite of chemical compounds should be present in the solar nebula as in flames, and the same reaction mechanisms should operate in the solar nebula as those which are observed for technical burning processes under low pressure conditions. 

For modelling the oxidation of the carbon dust and of the hydrocarbons released from the carbonaceous material we use the basic reaction mechanisms developed for the modelling of sooting flames \citep[e.g.][]{Mar96,Wan97,App00}. In our model calculation the H, C, and O chemistry is based on the model for combustion and soot formation of aliphatic fuels by \citet{Mar96}. Because of the low pressure prevailing in the accretion disk ($\lesssim100$ Pa, cf. Fig.~\ref{FigDiskProp}) the low-pressure limit of all pressure dependent reactions is used. Additionally He is included because this may be important for the few three-body reactions that may occur at the low pressures in accretion disks. The chemical reaction system used for this calculation consists of 146 species formed from H, C and O. It contains the OH-system for the formation/dissociation of water with eight species and 138 species with one to eighteen carbon atoms and the totally 1341 chemical reactions from \citet{Mar96} including the automatically generated reverse reactions where necessary.

\section{Sample calculation}
\label{SectGasChem}

We consider the oxidation of the mixture of carbonaceous materials assumed to have existed in the solar nebula. We add to the reaction network for the combustion chemistry of aliphatic fuels the components CH$_4$[v] to C$_{16}$H$_{10}$[r] as defined in Appendix~\ref{SectPyrCarb} and the equations for their pyrolysis. We use for their pyrolysis the rate coefficients as defined by Table~\ref{TabKinCarbComp}. The initial concentrations of the gas phase species are chosen as follows: all H is in H$_2$, one half of the C abundance in CO, and that part of the O not bound in minerals and CO is in H$_2$O. The concentration of all other gas phase species is set to zero. The corresponding element abundances are given in Table~\ref{TabSNAbu}. The initial concentrations of the components released at elevated temperature by pyrolysis are defined by Eq.~(\ref{InitialConcComp}). The complete set of reaction equations is solved numerically by a fully implicit Euler method. We start with an initial temperature of 300\,K and increase the temperature at a fixed rate of $10^{-2}\,\rm K\,yr^{-1}$ up to 1\,400\,K. The rate of temperature increase is chosen such that it corresponds approximately the temperature change during the accretion flow at about 1 AU distance in the disc model from Sect.~\ref{SectDiskMod}. A pressure of $p=10^{-4}$ bar is assumed. 

The resulting variation of the concentrations of the species are shown in Fig.~\ref{FigPyrolCarbMat} where the concentration of species relative to H-nuclei from the HO-chemistry and the species related to the CO-CO$_2$-chemistry are shown in Fig.~\ref{FigPyrolCarbMat}a while Fig.~\ref{FigPyrolCarbMat}b-k  shows the hydrocarbon species with 1 to 15 carbon atoms. The black lines show the concentrations of the different components of the solid carbonaceous material. These components disappear between $\mbox{$\sim350$}$ and $\sim500$ K by pyrolysis, except for the amorphous carbon which is oxidised at about 1\,200 K by reactions with OH radicals from the gas phase. 

The liberation of the pyrolysis products results in a rich gas phase chemistry of numerous hydrocarbon compounds from the whole size range from 1 to 15 carbon atoms. Even compounds with a higher number of C atoms are predicted to be formed if the network is extended to such compounds. This is in so far remarkable as in the oxygen-rich and hydrogen-rich environment of the solar nebula none of them would exist in abundance under chemical equilibrium conditions; CO would be the main carbon bearing gas phase species. The reason why such a rich zoo of hydrocarbons exists is that all possible oxidation reactions that finally convert hydrocarbons to CO involve rather high activation energies and do not become operative under the low pressure conditions prevailing in the solar nebula until the temperature exceeds about 1\,200 K. At lower temperatures the hydrocarbon chemistry cannot evolve into chemical equilibrium and hydrocarbons can coexist with water vapour in a metastable state \citep{Zol01}. Reactions between hydrocarbons and atomic H occur with significant rates such that all hydrocarbons with more than three carbon atoms are degraded to small hydrocarbons with 1 or 2 carbon atoms, mainly CH$_4$ and C$_2$H$_2$, before they disappear at temperatures above $\sim1\,200$ K by oxidation to CO. The final stage of the hydrocarbon combustion involves only compounds with one or two carbon atoms. This is a well known phenomenon in the combustion of organic fuels. More details of the gas phase chemistry will be discussed elsewhere. 
  
The important point is that the amorphous carbon component does not vanish from the dust component in the solar nebula before the OH concentration climbed to such a level that the rate of reaction (\ref{ReactSootOxi}) becomes sufficiently high. This requires temperatures of $T\gtrsim1\,200$ K. For planetesimals formed in disk regions with lower temperature they will incorporate the residual amorphous carbon at the onset of planetesimal formation.
 
These follow-up processes of the gasification of the carbonaceous material will be discussed in more detail elsewhere. 
 
\end{appendix}


\end{document}